\documentclass[11pt,a4wide]{article}

\newcommand{\remove}[1]{}
\usepackage[utf8]{inputenc}
\usepackage[english]{babel}

\usepackage{tikz,slashbox}
\usepackage{xspace}

\usepackage{amsmath,amsfonts,amssymb,epsfig,color}
\usepackage{stmaryrd}
\usepackage{amssymb}
\usepackage{amsfonts}
\usepackage{amsmath}
\usepackage{hyperref}
\usepackage{amsthm}
\usepackage{amssymb,amstext,hyperref}
\usepackage{cite}
\usepackage{graphics}
\usepackage{amsmath,dsfont}
\usepackage{mathrsfs}
\usepackage{fancyhdr}
\usepackage{verbatim}
\usepackage{boxedminipage}
\usepackage{colortbl}
\usepackage{algorithm}
\usepackage[noend]{algorithmic}

\usepackage[T1]{fontenc}
\usepackage{tikz}
\usepackage{graphicx}

\newtheorem{theorem}{Theorem}
\newtheorem{lemma}{Lemma}

\newcommand{\cupall}{\pmb{\pmb{\bigcup}}}

\newcommand{\Ostar}{\mathcal{O}^*}

\newcommand{\Acal}{\mathcal{A}}
\newcommand{\Bcal}{\mathcal{B}}

\newcommand{\Fcal}{\mathcal{F}}

\newcommand{\Ocal}{\mathcal{O}}
\newcommand{\Pcal}{\mathcal{P}}
\newcommand{\Qcal}{\mathcal{Q}}
\newcommand{\Rcal}{\mathcal{R}}

\newcommand{\Xcal}{\mathcal{X}}

\newcommand{\Nbb}{\mathbb{N}}

\newcommand{\ETH}{{\sf ETH}\xspace}

\newcommand{\Bound}[1]{{\bf #1}}
\newcommand{\bound}[1]{{\bf #1}}

\newcommand{\Chu}{\mu}

\newcommand{\bw}{{\sf bw}}

\newcommand{\psiG}[1]{\psi_{#1}}
\newcommand{\psiGi}[2]{\psi_{\Bound{#1}_{#2}}}

\newcommand{\degs}[2]{{\sf deg}_{#1}(#2)}

\theoremstyle{plain}
\newtheorem{proposition}[theorem]{Proposition}
\newtheorem{observation}{Observation}

\newcommand{\ig}[1]{\textcolor{red}{[Ig: #1]}}

\newcommand{\paraprobl}[5]
{
  \begin{flushleft}
    \fbox{
      \begin{minipage}{#5cm}
        \noindent {\textsc {#1}}\\
        {\bf Input:} #2\\
        {\bf Parameter:} #4\\
        {\bf Output:} #3
      \end{minipage}
    }
  \end{flushleft}
}

\newcommand{\sm}{\setminus}
\newcommand{\gm}{\setminus}

\newcommand{\tw}{{\sf{tw}}}

\newcommand{\es}{\emptyset}

\newcommand{\bs}{\backslash}

\newcommand{\intv}[1]{\left [ #1 \right ]}

\newcommand{\pretp}{\preceq_{\sf tm}}
\newcommand{\prem}{\preceq_{\sf m}}

\newcommand{\paw}{{\sf paw}\xspace}

\newcommand{\chair}{{\sf chair}\xspace}

\newcommand{\banner}{{\sf banner}\xspace}

\newcommand{\house}{{\sf px}\house}

\usepackage{aecompl}
\usepackage{color}
\definecolor{linkcol}{rgb}{0,0,0.8}
\definecolor{citecol}{rgb}{0.65,0,0}
\definecolor{titlecol}{rgb}{0.65,0,0}

\hypersetup {bookmarksopen=true, pdftitle="Hitting (topological) minors on
  bounded treewidth graphs", pdfauthor="Julien Baste - Ignasi Sau - Dimitrios M. Thilikos",
  pdfsubject="Hitting (topological) minors on
  bounded treewidth graphs", 
  pdfmenubar=true, 
  pdfhighlight=/O, 
  colorlinks=true, 
  pdfpagemode=None, 
  pdfpagelayout=DoublePage, 
  pdffitwindow=true, 
  linkcolor=linkcol, 
  citecolor=citecol, 
  urlcolor=linkcol 
}

\newcommand{\ex}{\textsf{ex}}

\definecolor{gray0}{gray}{0.875}
\definecolor{gray1}{gray}{0.775}
\definecolor{gray2}{gray}{0.75}

\newcommand\cuparrow{%
  \mathrel{\ooalign{\hss$\cup$\hss\cr%
      \kern0.3ex\raise0.7ex\hbox{\scalebox{0.7}{$\downarrow$}}}}}

\newcommand\bigcuparrow{%
  \mathrel{\ooalign{\hss$\bigcup$\hss\cr%
      \kern0.55ex\raise0.7ex\hbox{\scalebox{0.7}{$\downarrow$}}}}}

\newcommand{\pbtm}{\textsc{$\Fcal$-TM-Deletion} }
\newcommand{\pbm}{\textsc{$\Fcal$-M-Deletion} }

\begin{document}

\title{\vspace{-.5cm}Hitting minors on bounded treewidth graphs.\\I. General upper bounds\thanks{Emails of authors: \texttt{julien.baste@uni-ulm.de}, \texttt{ignasi.sau@lirmm.fr}, \texttt{sedthilk@thilikos.info}.\vspace{.2cm}\newline \indent \!\!\!\!\!The results of this article are permanently available at \texttt{https://arxiv.org/abs/1704.07284}. An extended abstract containing some of the results of this article appeared in the \emph{Proc. of the 12th International Symposium on Parameterized and Exact Computation  (\textbf{IPEC 2017})}~\cite{BasteST17}. Work supported by French projects DEMOGRAPH (ANR-16-CE40-0028) and ESIGMA (ANR-17-CE23-0010), and by the Deutsche Forschungsgemeinschaft (DFG, German Research Foundation) - 388217545.\newline}}

\author{\bigskip Julien Baste\thanks{LIRMM, Univ Montpellier, Montpellier, France.}~\thanks{Sorbonne Université, Laboratoire d'Informatique de Paris 6, LIP6, Paris, France.} \and
  Ignasi Sau\thanks{LIRMM, Univ Montpellier, CNRS, Montpellier, France.}
  \and
  Dimitrios  M. Thilikos$^{{\small \S}}$}

\date{\vspace{-1cm}}

\maketitle

\begin{abstract}
   \noindent For a finite collection of graphs ${\cal F}$, the \textsc{$\Fcal$-M-Deletion} problem consists in, given a graph $G$ and an integer $k$, deciding whether there exists $S \subseteq V(G)$ with $|S| \leq k$ such that $G \setminus S$ does not contain any of the graphs in ${\cal F}$ as a minor. We are interested in the parameterized complexity of \textsc{$\Fcal$-M-Deletion} when the parameter is the treewidth of $G$, denoted by $\tw$. Our objective is to determine, for a fixed ${\cal F}$, the smallest function $f_{{\cal F}}$ such that \textsc{$\Fcal$-M-Deletion} can be solved in time $f_{{\cal F}}(\tw) \cdot n^{\Ocal(1)}$ on $n$-vertex graphs. We prove that $f_{{\cal F}}(\tw) = 2^{2^{\Ocal (\tw \cdot\log \tw)}}$ for every collection ${\cal F}$, that $f_{{\cal F}}(\tw) = 2^{\Ocal (\tw \cdot\log \tw)}$ if ${\cal F}$ contains a planar graph, and that $f_{{\cal F}}(\tw) = 2^{\Ocal (\tw)}$ if in addition the input graph $G$ is planar or embedded in a surface.
We also consider the version of the problem where the graphs in ${\cal F}$ are forbidden as {\sl topological} minors, called  \textsc{$\Fcal$-TM-Deletion}. We prove similar results for this problem, except that in the last two algorithms, instead of requiring $\Fcal$ to contain a planar graph, we need it to contain a {\sl subcubic} planar graph.
This is the first of a series of articles on this topic.

\vspace{.5cm}

\vspace{.5cm}

\end{abstract}

\newpage

%

\section{Introduction}
\label{illusion}

Let ${\cal F}$ be a finite non-empty collection of non-empty graphs.  In the \textsc{$\Fcal$-M-Deletion} (resp. \textsc{$\Fcal$-TM-Deletion}) problem, we are given a graph $G$ and an integer $k$, and the objective is to decide whether there exists a set $S \subseteq V(G)$ with $|S| \leq k$ such that $G \setminus S$ does not contain any of the graphs in ${\cal F}$ as a minor (resp. topological minor). These problems belong to the wider family of {\sl graph modification problems} and  have a big expressive power, as instantiations of them correspond to several well-studied problems. For instance,  the cases ${\cal F}= \{K_2\}$, ${\cal F}= \{K_3\}$, and ${\cal F}= \{K_5,K_{3,3}\}$ of \textsc{$\Fcal$-M-Deletion} (or \textsc{$\Fcal$-TM-Deletion}) correspond to \textsc{Vertex Cover}, \textsc{Feedback Vertex Set}, and \textsc{Vertex Planarization}, respectively.

For the sake of readability, we use the notation \textsc{$\Fcal$-Deletion} in statements that apply to {\sl both} \textsc{$\Fcal$-M-Deletion} and  \textsc{$\Fcal$-TM-Deletion}. Note that if ${\cal F}$ contains a graph with at least one edge, then \textsc{$\Fcal$-Deletion} is {\sf NP}-hard by the classical classification result of Lewis and Yannakakis~\cite{LeYa80}.


We are interested in the parameterized complexity of \textsc{$\Fcal$-Deletion} when the parameter is the treewidth of the input graph (formally defined in~Section~\ref{banality}). Since the property of containing a graph as a (topological) minor can be expressed in Monadic Second Order logic (see~\cite{KimLPRRSS16line} for explicit formulas), by Courcelle's theorem~\cite{Courcelle90}, \textsc{$\Fcal$-Deletion} can be solved in time $\Ostar(f(\tw))$ on graphs with treewidth at most $\tw$, where $f$ is some computable function\footnote{The notation $\Ostar(\cdot)$  suppresses polynomial factors depending on the size of the input graph.}. Our objective is to determine, for a fixed collection ${\cal F}$, which is the {\sl smallest} such function $f$ that one can (asymptotically)  hope for, subject to reasonable complexity assumptions.

This line of research has attracted some attention  in the parameterized complexity community during the last years. For instance, \textsc{Vertex Cover} is easily solvable in time $\Ostar(2^{\Ocal(\tw)})$, called \emph{single-exponential}, by standard dynamic-programming techniques, and no algorithm with running time $2^{o(\tw)}\cdot n^{\Ocal(1)}$ exists, unless the Exponential Time Hypothesis (\ETH)\footnote{The \ETH implies that 3-\textsc{SAT} on $n$ variables cannot be solved in time $2^{o(n)}$; see~\cite{ImpagliazzoP01} for more details.} fails~\cite{ImpagliazzoP01}.

For \textsc{Feedback Vertex Set}, standard dynamic programming techniques give a running time of $\Ostar(2^{\Ocal(\tw \cdot \log \tw)})$, while the lower bound under the \ETH~\cite{ImpagliazzoP01} is again $2^{o(\tw)}\cdot n^{\Ocal(1)}$. This gap remained open for a while, until Cygan et al.~\cite{CyganNPPRW11} presented an optimal algorithm running in time $\Ostar(2^{\Ocal(\tw)})$, introducing  the celebrated \emph{Cut{\sl \&}Count}
technique. This article triggered several other techniques to obtain single-exponential algorithms for so-called \emph{connectivity problems} on graphs of bounded treewidth, mostly based on algebraic tools~\cite{BodlaenderCKN15,FominLPS16}.


Concerning \textsc{Vertex Planarization}, Jansen et al.~\cite{JansenLS14} presented an algorithm of time $\Ostar(2^{\Ocal(\tw \cdot \log \tw)})$ as a crucial subroutine in an {\sf FPT}-algorithm parameterized by $k$. Marcin Pilipczuk~\cite{Pili15} proved that this running time is {\sl optimal} under the \ETH, by using the framework introduced by Lokshtanov et al.~\cite{permuclique}  for proving superexponential lower bounds.

\medskip
\noindent
\textbf{Our results and techniques}. We present the following algorithms for \textsc{$\Fcal$-Deletion} parameterized by treewidth:


\begin{enumerate}
\item \label{unstable}For every $\mathcal{F}$, \textsc{$\Fcal$-Deletion} can be solved in time  $\Ostar\left(2^{2^{\Ocal(\tw \cdot \log \tw)}}\right)$.

\item \label{numerous}For every collection $\mathcal{F}$ containing at least one planar graph\footnote{In the conference version of this paper~\cite{BasteST17}, we further required that all the graphs in $\mathcal{F}$ are connected; here we improve the result by dropping this assumption.} (resp. subcubic planar graph), \textsc{$\Fcal$-M-Deletion} (resp. \textsc{$\Fcal$-TM-Deletion}) can be solved in time
  $\Ostar(2^{\Ocal(\tw  \cdot \log \tw)})$.

\item \label{vicinity}If the input graph $G$ is planar or, more generally, embedded in a surface of bounded genus, then \textsc{$\Fcal$-M-Deletion} (resp. \textsc{$\Fcal$-TM-Deletion} if additionally $\mathcal{F}$ contains a subcubic planar graph) can be solved in time
  $\Ostar(2^{\Ocal(\tw)})$ for every collection $\Fcal$.

\end{enumerate}

\noindent  Let us provide some ideas of the techniques that we use in our algorithms. Our first algorithm running in time $\Ostar\left(2^{2^{\Ocal(\tw \cdot \log \tw)}}\right)$ is not complicated and the running time is probably quite natural for anyone used to dynamic programming on graphs of bounded treewidth. The encoding that we use in the tables of the dynamic programming algorithm is based in the notion of \emph{folio}, which had been already used in previous work~\cite{AdlerDFST11,RobertsonS95b}. Informally speaking, the folio of a graph contains all the ``partial models'' of the graphs in $\Fcal$ that survive after the removal of a partial solution (see Section~\ref{banality} for the formal definition). It is easy to see that keeping track of the folios is enough to solve both problems and that the size of a folio is bounded by $2^{\Ocal(\tw \cdot \log \tw)}$ (cf. Lemma~\ref{humanism}), hence the number of distinct folios is at most $2^{2^{\Ocal(\tw \cdot \log \tw)}}$, yielding the claimed running time.

The algorithm running in time $\Ostar(2^{\Ocal(\tw  \cdot \log \tw)})$, when $\Fcal$ contains a (subcubic) planar graph, uses the machinery of boundaried graphs, equivalence relations, and representatives originating in the seminal work of Bodlaender et al.~\cite{BodlaenderFLPST16} (see also~\cite{BodlaenderV01redu}) and  subsequently used, for instance, in~\cite{GarneroPST15,F.V.Fomin:2010oq,KimLPRRSS16line}. The main conceptual difference with respect to the algorithm discussed above is that the encoding in the tables of the dynamic programming algorithm, which we use to construct the partial solutions, is not based on the notion of folio anymore, but on the notion of \emph{representative} of an appropriately defined equivalence relation. Intuitively, such a representative corresponds to a possible behavior of a boundaried graph (associated with the subgraph of the input graph $G$ rooted at a bag of a tree decomposition) with respect to the eventual occurrences of graphs in $\Fcal$ when ``gluing'' another unknown boundaried graph to it, corresponding to the subgraph of $G$ that has not been explored yet (again, see Section~\ref{banality} for the formal definition). The fact that $\Fcal$ contains a (subcubic) planar graph is essential in order to bound the treewidth of the resulting graph after deleting a partial solution (cf. Lemma~\ref{granting}) and this is crucially used in order to bound the number of representatives (cf. Proposition~\ref{touching}). For technical reasons, in all our algorithms we use {\sl branch} decompositions instead of tree decompositions, whose associated widths are equivalent from a parametric point of view~\cite{RobertsonS91}.

The algorithm running in time $\Ostar\left(2^{\Ocal(\tw)}\right)$, when the input graph $G$ is planar,  exploits sphere-cut decompositions~\cite{SeymourT94,DornPBF10}, a special type of branch decompositions of planar graphs with nice topological properties. We prove that, if we use sphere-cut decompositions and we apply essentially the same dynamic programming algorithm discussed above, the number of representatives can be upper-bounded by the number of (unlabeled) planar graphs on $\Ocal(\tw)$ vertices, which are $2^{\Ocal(\tw)}$ many~\cite{Tut62}. With some more technical details, we extend this single-exponential algorithm to graphs embedded in surfaces by using
surface-cut decompositions, introduced by Ru\'e et al.~\cite{RueST14}.

We present these algorithms for the topological minor version, and then it is easy to adapt them to the minor version within the claimed running time (cf. Lemma~\ref{sickness}).

\medskip
\noindent
\textbf{Results in other articles of the series}. In the second article of this series~\cite{monster2}, we show that if $\Fcal \in \{\{P_3\}, \{P_4\}, \{K_{1,i}\},\{C_4\}, \{\paw\}, \{\chair\}, \{\banner\}\}$, then \pbtm 
can be solved in single-exponential time. Note that all these graphs have maximum degree at most three, except $K_{1,i}$ for $i \geq 4$, and therefore the corresponding algorithms also apply to the \pbm problem. In the third article of this series~\cite{monster3}, we focus on lower bounds under the \ETH. Namely, we prove that for any collection $\Fcal$ containing only connected graphs of size at least two, \textsc{$\Fcal$-Deletion} cannot be solved in time ${2^{o(\tw)} \cdot n^{\Ocal(1)}}$, even if the input graph $G$ is planar, and we also provide superexponential lower bounds for a number of collections $\Fcal$. In particular, we prove a lower bound of ${2^{o(\tw \cdot \log \tw)}\cdot n^{\Ocal(1)}}$ when $\Fcal$ contains a single graph that is either $P_5$ or is not a minor of the $\banner$ (that is,
the graph consisting of a $C_4$ plus a pendent edge), with the exception of $K_{1,i}$ for the topological minor version. These lower bounds, together with the ad hoc single-exponential  algorithms given in~\cite{monster3} and the algorithm described in item~\ref{numerous} above, cover all the cases of \textsc{$\Fcal$-M-Deletion} where $\Fcal$ consists of a single connected planar graph $H$, yielding a tight dichotomy in terms of~$H$. In the fourth article of this series~\cite{BasteST20-SODA} (whose full version is~\cite{SODA-arXiv}), we presented an algorithm for \textsc{$\Fcal$-M-Deletion} in time $\Ostar(2^{O(\tw \cdot \log \tw)})$ for {\sl any} collection $\Fcal$, yielding together with the lower bounds in~\cite{monster3} a dichotomy for \textsc{$\Fcal$-M-Deletion} where $\Fcal$ consists of a single connected (not necessarily planar) graph $H$. 

\medskip
\noindent
\textbf{Organization of the paper}. In Section~\ref{banality} we give some preliminaries, and in  Section~\ref{enclaves} we formally state the results of this article. In Section~\ref{dispense} we introduce the formalism of boundaried graphs and their equivalence classes, and prove several technical lemmas.
In Section~\ref{dictates} we define branch decompositions of boundaried graphs and prove some basic properties. We prove the result of item~\ref{unstable} in Section~\ref{redeemed}.
In Section~\ref{expected} we provide improved bounds on the sets of representatives in the case where ${\cal F}$ contains a planar (subcubic) graph, and we use this result in Section~\ref{dramatic} to prove  the result of item~\ref{numerous}. Finally, we prove  the result of item~\ref{vicinity} in Section~\ref{features} (for planar graphs) and Section~\ref{crinkled} (for bounded-genus graphs).
We conclude the article in Section~\ref{upheaval} with some open questions for further research.

\section{Preliminaries}
\label{banality}

In this section we provide some preliminaries to be used in the following sections.


\bigskip
\noindent
\textbf{Sets, integers, and functions.}
We denote by $\Nbb$ the set of every non-negative integer and we
set $\Nbb^+=\Nbb\setminus\{0\}$.
Given two integers $p$ and $q$, the set $\intv{p,q}$
refers to the set of every integer $r$ such that $p \leq r \leq q$.
Moreover, for each integer $p \geq 1$, we set $\Nbb_{\geq p} = \Bbb{N}\setminus\intv{0,p-1}$.


We use $\emptyset$ to denote the empty set and
$\varnothing$ to denote the empty function, i.e., the unique subset of $\emptyset\times\emptyset$.
Given a function $f:A\to B$ and a set $S$, we define $f|_{S}=\{(x,f(x))\mid x\in S\cap A\}$.
Moreover if $S \subseteq A$, we set $f(S) = \bigcup_{s \in S} \{f(s)\}$.
Given a set $S$, we denote by ${S \choose 2}$ the set containing
every subset of $S$ that has cardinality two. We also denote by $2^{S}$ the set of all the subsets of $S$.
If ${\cal S}$ is a collection of objects where the operation $\cup$ is defined, then we  denote $\cupall{\cal S}=\bigcup_{X\in{\cal S}}X$.

Let $p\in\Bbb{N}$ with $p\geq 2$, let $f:\Bbb{N}^{p}\rightarrow\Bbb{N}$, and let  $g:\Bbb{N}^{p-1}\rightarrow\Bbb{N}$.
We say that $f(x_{1},\ldots,x_{p})=\Ocal_{x_p}(g(x_{1},\ldots,x_{p-1}))$ if there is a function   $h:\Bbb{N}\rightarrow\Bbb{N}$ such that  \\ $f(x_{1},\ldots,x_{p})=\Ocal(h(x_p)\cdot g(x_{1},\ldots,x_{p-1}))$.

\bigskip
\noindent
\textbf{Graphs}.  All the graphs that we consider in this paper are undirected, finite, and without  loops or multiple edges (except for the graph $\theta_s$, sometimes called \emph{pumpkin} in the literature~\cite{JoretPSST14}). We use standard graph-theoretic notation, and we refer the reader to~\cite{Die10} for any undefined terminology. Given a graph $G$, we denote by $V(G)$ the set of vertices of  $G$ and by $E(G)$ the set of the edges of $G$.
We call $|V(G)|$ {\em the size} of $G$. A graph is {\em the empty} graph if its size is zero.
We also denote by $L(G)$ the set of the vertices of $G$ that have degree exactly
1.
If $G$ is a tree (i.e., a connected acyclic graph) then $L(G)$ is the set of the {\em leaves}
of $G$.
A {\em vertex labeling} of $G$ is some injection $\rho: V(G)\to\Bbb{N}^{+}$. 
{Given a vertex $v \in V(G)$, we define the \emph{neighborhood} of
  $v$ as $N_G(v) = \{u \mid u \in V(G), \{u,v\} \in E(G)\}$ and the \emph{closed neighborhood}
  of $v$ as $N_G[v] = N_G(v) \cup \{v\}$.} If $X\subseteq V(G)$, then we write $N_{G}(X)=(\bigcup_{v\in X}N_{G}(v))\setminus X$.
The {\em degree} of a vertex $v$ in $G$ is defined as $\degs{G}{v}=|N_{G}(v)|$. A graph is called {\em subcubic}
if all its vertices have degree at most three.

A \emph{subgraph} $H = (V_H,E_H)$ of a graph $G=(V,E)$ is a graph such that $V_H \subseteq V(G)$ and $E_H \subseteq E(G) \cap {V(H) \choose 2}$.
If $S \subseteq V(G)$, the subgraph of $G$ \emph{induced by} $S$, denoted $G[S]$, is the graph $(S, E(G) \cap {S \choose 2})$.
We also define $G \gm S$ to be the subgraph of $G$ induced by $V(G) \sm S$.
If $S \subseteq E(G)$, we denote by $G \gm S$ the graph $(V(G), E(G) \sm S)$.
%

If $s,t\in V(G)$, an {\em $(s,t)$-path} of $G$
is any connected subgraph $P$ of $G$ with maximum degree two and where $s,t\in L(P)$.
We finally denote by ${\cal P}(G)$ the set of all paths of $G$.
{Given $P \in \Pcal(G)$, we say that $v \in V(P)$ is an \emph{internal vertex} of $P$ if $\degs{P}{v} = 2$.}
{Given an integer $i$ and a graph $G$, we say that $G$ is $i$-connected if for each $\{u,v\} \in {V(G) \choose 2}$,
  there exists a set $\Qcal \subseteq \Pcal(G)$ of $(u,v)$-paths of $G$ such that
  $|\Qcal| = i$ and
  for each $P_1,P_2 \in \Qcal$ such that $P_1 \not = P_2$, $V(P_1) \cap V(P_2) = \{u,v\}$.} We denote by $K_{r}$,  $P_{r}$, and $C_{r}$, the complete graph, the path, and the cycle  on $r$ vertices, respectively.

\bigskip
\noindent
\textbf{Minors and topological minors.}
Given two graphs $H$ and $G$ and two functions $\phi: V(H)\to V(G)$ and $\sigma: E(H)\to{\cal P}(G)$, we say that $(\phi,\sigma)$ is {\em a topological minor model of $H$ in $G$}
if
\begin{itemize}
\item for every $\{x,y\}\in E(H),$ $\sigma(\{x,y\})$ is an $(\phi(x),\phi(y))$-path 
  in $G$,
\item $\phi$ is an injective function, and
\item if $P_{1},P_{2}$ are two distinct paths in $\sigma(E(H))$, then none of the internal vertices of $P_{1}$
  is a vertex of $P_{2}$.
\end{itemize}

The {\em branch} vertices of
$(\phi,\sigma)$ are the vertices in  $\phi(V(H))$, while the {\em subdivision} vertices of $(\phi,\sigma)$ are the internal vertices of the paths in $\sigma(E(H))$.

We say that $G$ contains $H$ as a \emph{topological minor}, denoted by $H\preceq_{\sf tm} G$, if there
is a topological minor model  $(\phi,\sigma)$ of $H$ in $G$.


Given two graphs $H$ and $G$ and a function $\phi: V(H)\to 2^{V(G)}$, we say that $\phi$ is {\em a minor model of $H$ in $G$}
if
\begin{itemize}
\item for every $x \in V(H)$, $G[\phi(x)]$ is a connected non-empty graph,
\item if $x,y$ are two distinct vertices of $H$, then $\phi(x) \cap \phi(y) = \emptyset$, and
\item for every $\{x,y\} \in E(H)$, there exist  $x' \in \phi(x)$ and $y' \in \phi(y)$ such that $\{x',y'\} \in E(G)$.
\end{itemize}

We say that $G$ contains $H$ as a \emph{minor}, denoted by $H\prem G$, if there
is a minor model  $\phi$ of $H$ in $G$.

Let $H$ be a graph. We define the set of graphs ${\sf tpm}(H)$ as follows: among all the graphs containing $H$ as a minor,
we consider only those that are minimal with respect to the topological minor relation. The following two observations follow easily from the above definitions.


\begin{observation}
  \label{stuffing}
  There is a function $f_{1}:\Bbb{N}\to\Bbb{N}$
  such that for every $h$-vertex graph $H$, every graph in ${\sf tpm}(H)$ has at most $f_{1}(h)$ vertices.
\end{observation}

An explicit function $f_1$ as in Observation~\ref{stuffing} can be obtained by replacing every vertex $v \in V(H)$ with $\degs{H}{v}\geq 4$ by a tree whose leaf set is $N_H(v)$ and with no internal vertices of degree two. Clearly, for each vertex $v \in V(H)$, the number of such trees (each yielding a distinct graph in ${\sf tpm}(H)$) and their size depend only on $\degs{H}{v}$.

\begin{observation}
  \label{presents}
  Given two graphs $H$ and $G$, $H$ is a minor of $G$ if and only if there exists some graph in ${\sf tpm}(H)$ that is a topological minor of $G$.
\end{observation}

\bigskip
\noindent
\textbf{Graph collections.}
Let ${\cal F}$  be a collection of graphs. From now on instead of ``collection of graphs''  we use the shortcut ``collection''.
If ${\cal F}$ is a collection that is finite, non-empty, and all its graphs are  non-empty, then we say that ${\cal F}$
is a {\em proper collection}. For any proper collection ${\cal F}$, we  define
${\sf size}({\cal F})=\max\{|V(H)|\mid H\in \cal F\}$.
Note that if the size of ${\cal F}$ is bounded, then
the size of the graphs in ${\cal F}$ is also bounded.
We  say that $\Fcal$ is an {\em (topological) minor antichain}
if no two of its elements are comparable via the  (topological)  minor relation.

Let $\Fcal$ be a proper collection. We extend the (topological) minor relation to $\Fcal$ such that, given a graph $G$,
$\Fcal \preceq_{\sf tm} G$ (resp. $\Fcal \preceq_{\sf m} G$) if and only if there exists a graph $H \in \Fcal$ such that $H \preceq_{\sf tm} G$ (resp. $H \preceq_{\sf m} G$).
We also denote $\ex_{\sf tm}(\Fcal)=\{G\mid \Fcal\npreceq_{\sf tm} G \}$, i.e.,
$\ex_{\sf tm}(\Fcal)$ is the class of graphs that do not contain any graph in $\Fcal$ as a topological minor.
The set $\ex_{\sf m}(\Fcal)$ is defined analogously.

\bigskip
\noindent\textbf{Tree decompositions.} A \emph{tree decomposition} of a graph $G$ is a pair ${\cal D}=(T,{\cal X})$, where $T$ is a tree
and ${\cal X}=\{X_{t}\mid t\in V(T)\}$ is a collection of subsets of $V(G)$
such that:
\begin{itemize}
\item $\bigcup_{t \in V(T)} X_t = V(G)$,
\item for every edge $\{u,v\} \in E$, there is a $t \in V(T)$ such that $\{u, v\} \subseteq X_t$, and
\item for each $\{x,y,z\} \subseteq V(T)$ such that $z$ lies on the unique path between $x$ and $y$ in $T$,  $X_x \cap X_y \subseteq X_z$.
\end{itemize}
We call the vertices of $T$ {\em nodes} of ${\cal D}$ and the sets in ${\cal X}$ {\em bags} of ${\cal D}$. The \emph{width} of a  tree decomposition ${\cal D}=(T,{\cal X})$ is $\max_{t \in V(T)} |X_t| - 1$.
The \emph{treewidth} of a graph $G$, denoted by $\tw(G)$, is the smallest integer $w$ such that there exists a tree decomposition of $G$ of width at most $w$.
For each $t \in V(T)$, we denote by $E_t$ the set $E(G[X_t])$.


\bigskip
\noindent\textbf{Parameterized complexity.} We refer the reader to~\cite{DF13,CyganFKLMPPS15} for basic background on parameterized complexity, and we recall here only some very basic definitions.
A \emph{parameterized problem} is a language $L \subseteq \Sigma^* \times \mathbb{N}$.  For an instance $I=(x,k) \in \Sigma^* \times \mathbb{N}$, $k$ is called the \emph{parameter}. 
A parameterized problem is \emph{fixed-parameter tractable} ({\sf FPT}) if there exists an algorithm $\Acal$, a computable function $f$, and a constant $c$ such that given an instance $I=(x,k)$,
$\Acal$ (called an {\sf FPT} \emph{algorithm}) correctly decides whether $I \in L$ in time bounded by $f(k) \cdot |I|^c$.


\bigskip
\noindent
\textbf{Definition of the problems.} Let ${\cal F}$ be a proper collection.
We define the parameter ${\bf tm}_{\cal F}$ as the function that maps graphs to non-negative integers as follows:
\begin{eqnarray}
  \label{overlaid}
  {\bf tm}_{\cal F}(G) & = & \min\{|S|\mid  S\subseteq V(G)\wedge G\setminus S\in{\sf ex}_{\sf tm}({\cal F})\}.
\end{eqnarray}
The parameter ${\bf m}_{\cal F}$ is defined analogously.
The main objective of this paper is to study the problem of computing the parameters
${\bf tm}_{\cal F}$ and ${\bf m}_{\cal F}$  for graphs of bounded treewidth under several instantiations of
the collection ${\cal F}$. The corresponding decision problems are formally defined as follows.
\medskip\medskip\medskip

\hspace{-.6cm}\begin{minipage}{6.7cm}
  \paraprobl{\textsc{$\Fcal$-TM-Deletion}}
  {A graph $G$ and an integer $k\in \Nbb$.}
  {Is ${\bf tm}_\Fcal(G)\leq k$?}
  {The treewidth of $G$.}{6.7}
\end{minipage}~~~~~
\begin{minipage}{6.7cm}
  \paraprobl{\textsc{$\Fcal$-M-Deletion}}
  {A graph $G$ and an integer $k\in \Nbb$.}
  {Is ${\bf m}_\Fcal(G)\leq k$?}
  {The treewidth of $G$.}{6.7}
\end{minipage}
\medskip\medskip

Note that in both the above problems,
we can always assume that ${\cal F}$ is an antichain
with respect to the considered relation.  Indeed, this is the case because if ${\cal F}$
contains two graphs $H_{1}$ and $H_{2}$ where $H_{1}\preceq_{\sf tm} H_{2}$, then
${\bf tm}_{\cal F}(G)={\bf tm}_{{\cal F}'}(G)$ where ${\cal F}'={\cal F}\setminus\{H_{2}\}$ (similarly
for the minor relation).

Throughout the article, we let $n$ and $\tw$ be the number of vertices and the treewidth of the input graph of the considered problem, respectively.






%

%
%

\section{Formal statement of the results}
\label{enclaves}


%
%





The purpose of the remainder of the article is to prove the following results.




\begin{theorem}
  \label{perishes}
  If ${\cal F}$ is a proper collection, where $d={\sf size}({\cal F}),$ then
  there exists an algorithm that solves \pbtm
  in $2^{2^{\Ocal_{d}(\tw \cdot \log \tw)}}\cdot n$ steps.
\end{theorem}

\begin{theorem}
  \label{whipping}
  If ${\cal F}$ is a proper collection, where $d={\sf size}({\cal F}),$ then
  there exists an algorithm that solves \pbm
  in $2^{2^{\Ocal_{d}(\tw\cdot \log \tw)}}\cdot n$ steps.
\end{theorem}

\begin{theorem}
  \label{oblivion}
  If ${\cal F}$ is a proper collection containing a subcubic planar graph, where $d={\sf size}({\cal F}),$ then
  there exists an algorithm that solves \pbtm
  in ${2^{\Ocal_{d}(\tw \cdot \log \tw)}}\cdot n$ steps.
\end{theorem}


\begin{theorem}
  \label{headings}
  If ${\cal F}$ is a proper collection containing a planar graph, where $d={\sf size}({\cal F}),$ then
  there exists an algorithm that solves \pbm
  in ${2^{\Ocal_{d}(\tw\cdot \log \tw)}}\cdot n$ steps.
\end{theorem}

\begin{theorem}
  \label{deprives}
  If ${\cal F}$ is a proper collection containing a subcubic planar graph, where $d={\sf size}({\cal F}),$ then
  there exists an algorithm that solves \pbtm
  on planar graphs
  in ${2^{\Ocal_{d}(\tw)}}\cdot n+\Ocal(n^3)$ steps.
\end{theorem}

\begin{theorem}
  \label{cockneys}
  If ${\cal F}$ is a proper collection, where $d={\sf size}({\cal F}),$ then
  there exists an algorithm that solves \pbm
  on planar graphs
  in ${2^{\Ocal_{d}(\tw)}}\cdot n+\Ocal(n^3)$ steps.
\end{theorem}


It is worth mentioning that the lower bounds given in~\cite{monster3} imply that the algorithms of Theorems~\ref{deprives} and~\ref{cockneys} are asymptotically {\sl tight} under the \ETH. Note also that in both theorems one can assume that ${\cal F}$ contains only planar graphs, as an input planar graph $G$ does not contain any nonplanar graph as a (topological) minor. In Section~\ref{crinkled}, we present a generalization of Theorems~\ref{deprives} and~\ref{cockneys} to input graphs embedded in surfaces of arbitrary genus.

\medskip

The following lemma is a direct consequence of Observation~\ref{presents}.

\begin{lemma}
  \label{sickness}
  Let ${\cal F}$ be a proper collection. Then, for every graph $G,$
  it holds that  ${\bf m}_{\cal F}(G)={\bf tm}_{\cal F'}(G)$ where
  ${\cal F}'=\bigcup_{F\in{\cal F}}{\sf tpm}(F)$.
\end{lemma}


It is easy to see that for every (planar) graph $F,$ the set ${\sf tpm}(F)$ contains a subcubic (planar) graph; see the paragraph after Observation~\ref{stuffing}. Combining this observation with Lemma~\ref{sickness} and Observation~\ref{stuffing}, Theorems~\ref{whipping},~\ref{headings}, and~\ref{cockneys} follow directly from Theorems~\ref{perishes},~\ref{oblivion}, and~\ref{deprives}, respectively.
Therefore, the following sections are devoted to the proofs of Theorems~\ref{perishes},~\ref{oblivion}, and~\ref{deprives}.

%

%

\section{Boundaried graphs and their equivalence classes}
\label{dispense}

Many of the following definitions were introduced in~\cite{BodlaenderFLPST16,F.V.Fomin:2010oq} (see also~\cite{GarneroPST15,KimLPRRSS16line}).

\bigskip

\noindent
\textbf{Boundaried graphs.} Let $t\in\Bbb{N}$.
A \emph{$t$-boundaried graph} is a triple $\bound{G} = (G,R,\lambda)$ where $G$ is a graph, $R \subseteq V(G),$ $|R| = t,$ and
$\lambda : R \rightarrow \Nbb^+$ is an
injective function.
We call $R$ the {\em boundary} of ${\bf G}$ and we call the  vertices
of $R$ {\em the boundary vertices} of ${\bf G}$. We also
call $G$ {\em the underlying graph} of ${\bf G}$.
Moreover, we call $t=|R|$ the {\em boundary
  size} of ${\bf G}$ and we define the {\em label set} of $\bound{G}$  as $\Lambda({\bf G})=\lambda(R)$.
We also say that $\bound{G}$ is a \emph{boundaried graph} if there exists an integer $t$ such that $\bound{G}$ is an $t$-boundaried graph. We say that a boundaried graph ${\bf G}$ is {\em consecutive}
if $\Lambda({\bf G})=\intv{1,|R|}$.
We define the {\em size} of $\bound{G} = (G,R,\lambda),$ as $|V(G)|$ and we use the notation $V({\bf G})$ and  $E({\bf G})$ for   $V({G})$ and $E({G}),$ respectively.
If $S \subseteq V(G),$ we define $\bound{G}' = \bound{G} \setminus S$
such that $\bound{G}' = (G', R', \lambda'),$ $G' = G \setminus  S,$ $R'= R \sm S,$ and $\lambda'=\lambda|_{R'}$.
We define  $\Bcal^{(t)}$ as the set of all $t$-boundaried graphs.
We also use the notation ${\bf B}_{\varnothing}=((\emptyset,\{\emptyset\}),\emptyset,\varnothing)$
to denote the (unique) $0$-boundaried {\em empty boundaried graph}.

Given a $t$-boundaried graph $\bound{G}=(G,R,\lambda),$ we define $\psiG{\bound{G}} : R \rightarrow \intv{1,t}$ such that
for each $v \in R,$ $\psiG{\bound{G}}(v) =|\{u \in R \mid \lambda(u) \leq \lambda(v)\}| $.
Note that, as $\lambda$ is an injective function,  $\psiG{\bound{G}}$ is a bijection
and, given a boundary vertex $v$ of ${\bf G},$ we call $\psi_{\bf G}(v)$ the {\em index} of $v$.

Let $t\in\Bbb{N}$.
We  say that two $t$-boundaried graphs ${\bf G}_{1}=(G_{1},R_{1},\lambda_{1})$ and ${\bf G}_{2}=(G_{2},R_{2},\lambda_{2})$ are \emph{isomorphic}
if there is a bijection $\sigma: V({\bf G}_1)\rightarrow V({\bf G}_{2})$
that is an isomorphism $\sigma: V(G_{1})\to V(G_{2})$ from $G_1$ to $G_{2}$ and additionally
$\psi_{{\bf G}_{1}}^{-1}\circ \psi_{{\bf G}_{2}}\subseteq \sigma,$ i.e., $\sigma$ sends the boundary vertices of ${\bf G}_{1}$ to equally-indexed boundary vertices of ${\bf G}_{2}$.
We  say that  ${\bf G}_1$ and ${\bf G}_{2}$
are {\em boundary-isomorphic} if $\psi_{{\bf G}_{1}}^{-1}\circ \psi_{{\bf G}_{2}}$  is an isomorphism from $G_{1}[R_{1}]$ to $G_{2}[R_{2}]$ and we denote this fact by ${\bf G}_{1}\sim {\bf G}_{2}$. It is easy to make the following observation.

\begin{observation}
  \label{elegance}
  For every $t\in\Bbb{N},$ if ${\cal S}$ is a
  collection of $t$-boundaried graphs where $|{\cal S}|>2^{t \choose 2},$
  then ${\cal S}$ contains at least two
  boundary-isomorphic graphs.
\end{observation}

\bigskip
\noindent
\textbf{Topological minors of boundaried graphs.}
Let  $\bound{G}_1=(G_1,R_1,\lambda_1)$ and $\bound{G}_2=(G_2,R_2,\lambda_2)$ be
two boundaried graphs. We say that ${\bf G}_{1}$ is a {\em topological minor} of ${\bf G}_{2}$ if there is a topological minor model $(\phi,\sigma)$ of $G_1$ in $G_2$
such that
\begin{itemize}
\item $\psi_{{\bf G}_{1}}=\psi_{{\bf G}_{2}}\circ\phi|_{R_{1}} ,$ i.e.,
  the vertices of $R_{1}$ are mapped via $\phi$ to equally indexed vertices  of $R_{2}$  and
\item none of the vertices in $R_{2}\setminus \phi(R_{1})$
  is a subdivision vertex of $(\phi,\sigma)$.
\end{itemize}
We extend the definition of $\preceq_{\sf tm}$ so that ${\bf G}_{1}\preceq_{\sf tm}{\bf G}_{2}$ denotes that
${\bf G}_{1}$ is a topological minor of ${\bf G}_{2}$ and, given a ${\cal X}\subseteq {\cal B}^{(t)}$,
${\cal X}\preceq_{\sf tm}{\bf G}_{2}$ denotes that at least one of the $t$-boundaried graphs in
${\cal X}$ is a topological minor of the $t$-boundaried graph ${\bf G}_{2}$.


\bigskip
\noindent
\textbf{Operations on boundaried graphs.}
Let  $\bound{G}_1=(G_1,R_1,\lambda_1)$ and $\bound{G}_2=(G_2,R_2,\lambda_2)$ be
two $t$-boundaried graphs.
We define the \emph{gluing operation} $\oplus$ such that $(G_1, R_1, \lambda_1) \oplus (G_2, R_2, \lambda_2)$ is
the graph $G$ obtained by taking the disjoint union of $G_1$ and $G_2$ and then, for each $i \in \intv{1,t},$ identifying the vertex $\psiGi{G}{1}^{-1}(i)$ and the vertex $\psiGi{G}{2}^{-1}(i)$. Keep in mind that ${\bf G}_{1}\oplus{\bf G}_{2}$ is a graph and not a boundaried graph. Moreover, the operation $\oplus$ requires both boundaried graphs to have boundaries of the same size.

Let ${\bf G}=(G,R,\lambda)$ be a $t$-boundaried graph and let $I\subseteq\Bbb{N}$.
We use the notation ${\bf G}|_I=(G,\lambda^{-1}(I),\lambda|_{\lambda^{-1}(I)}),$ i.e., we do not include in the boundary anymore the vertices that are not indexed by numbers in $I$.
Clearly, ${\bf G}|_{I}$ is a $t'$-boundaried graph where $t'=|I\cap \Lambda(\bound{G})|$.

Let  $\bound{G}_1=(G_1,R_1,\lambda_1)$ and $\bound{G}_2=(G_2,R_2,\lambda_2)$ be
two boundaried graphs. Let also $I=\lambda_{1}(R_{1})\cap \lambda_{2}(R_{2})$
and let $t=|R_{1}|+|R_{2}|-|I|$.
We define the \emph{merging operation} $\odot$ such that $(G_1,R_1,\lambda_1) \odot (G_2,R_2,\lambda_2)$ is
the $t$-boundaried graph $G=(G,R,\lambda)$  where $G$ is obtained by taking
the disjoint union of $G_{1}$ and $G_{2}$ and then for each $i\in I$ identify
the vertex $\lambda^{-1}_1(i)$ with the vertex  $\lambda^{-1}_2(i)$. Similarly,
$R$ is obtained by $R_{1}\cup R_{2}$ after applying the same identifications
to pairs of vertices in $R_{1}$ and $R_{2}$. Finally, $\lambda=\lambda_{1}'\cup \lambda_{2}'$
where, for $j\in\intv{1,2},$ $\lambda_{j}'$ is obtained from $\lambda_{j}$ after replacing each $(x,i)\in\lambda_j$
(for some  $i\in I$) by $(x_{\rm new},i),$ where $x_{\rm new}$ is the result of the identification
of $\lambda_{1}^{-1}(i)$ and $\lambda_{2}^{-1}(i)$.
Observe that ${\bf G}_{1}\odot{\bf G}_{2}$ is a boundaried graph
and that the operation $\odot$ does not require input boundaried graphs to have  boundaries of the same size.

Let ${\bf G}=(G,R,\lambda)$ be a consecutive $t$-boundaried graph and let $I\subseteq \Bbb{N}$ be such that $|I| = t$.
We define ${\bf G}=(G,R,\lambda)\diamond I$ as the unique $t$-boundaried graph
 ${\bf G}'=(G,R,\lambda')$ where $\lambda': R \to I$ is a bijection and  $\psi_{{\bf G}'}=\lambda$.

\bigskip
\noindent
\textbf{Equivalence relations.}
Let $\Fcal$ be a proper collection and let $t$ be a non-negative integer.
We define an equivalence relation $\equiv^{(\Fcal,t)}$ on $t$-boundaried graphs as follows: Given two $t$-boundaried graphs $\bound{G}_1$ and $\bound{G}_2,$ we write
$\bound{G}_1 \equiv^{(\Fcal,t)}\bound{G}_2$ to denote  that
\vspace{-4mm}

\begin{eqnarray*}
  \forall \bound{G}\in\Bcal^{(t)}& \Fcal \pretp \bound{G}\oplus \bound{G}_1 \iff \Fcal \pretp\bound{G}\oplus \bound{G}_2.\label{ftigkeit}
\end{eqnarray*}
It is easy to verify that $\equiv^{(\Fcal,t)}$ is an equivalence relation.
We set up a {\em set of
  representatives} $\Rcal^{({\cal F},t)}$  as a set containing, for each equivalence
class ${\cal C}$ of $\equiv^{(\Fcal,t)},$ some consecutive $t$-boundaried graph in ${\cal C}$  with
minimum number of edges and, among those with minimum number of edges, with minimum number of vertices
(if there are more than one such graphs, pick one arbitrarily).
Given a $t$-boundaried graph ${\bf G}$ we denote by ${\sf rep}_{\cal F}({\bf G})$ the $t$-boundaried graph
${\bf B}\in\Rcal^{({\cal F},t)}$ where ${\bf B}\equiv^{(\Fcal,t)}{\bf G}$ and we call ${\bf B}$ the {\em ${\cal F}$-representative of ${\bf G}$}.  Clearly, ${\sf rep}_{\cal F}({\bf B})={\bf B}$.

Note that if ${\bf B}=(B,R,\lambda)$ is a $t$-boundaried graph
and ${\cal F}\pretp B,$ then ${\sf rep}_{\cal F}({\bf B})$
is, by definition, a consecutive $t$-boundaried graph whose underlying graph is a graph $H \in {\cal F}$ with minimum number of edges (and out of those, with minimum number of vertices), possibly completed with $t - |V(H)|$ isolated vertices in the case where $|V(H)| < t$.
%
%
We
denote this graph by ${\bf F}^{({\cal F},t)}$ (if there are many possible choices, just pick one arbitrarily).
Note also that the underlying graph of every boundaried graph in ${\cal R}^{({\cal F},t)}\setminus\{{\bf F}^{({\cal F},t)}\}$ belongs to ${\sf ex}_{\sf tm}({\cal F})$.

%
%

We need the following three  lemmas. The first one is a direct consequence of the definitions of the equivalence relation $\equiv^{(\Fcal,t)}$ and the set of representatives  $\Rcal^{({\cal F},t)}$.
%

\begin{lemma}
  \label{proceeds}
  Let ${\cal F}$ be a proper collection and let $t\in\Bbb{N}$.
  Let also ${\bf B}_{1}$ and ${\bf B}_{2}$ be $t$-boundaried graphs.
  Then ${\bf B}_{1}\equiv^{(\Fcal,t)}{\bf B}_{2}$ if and only if
  $\forall \bound{G}\in\Rcal^{({\cal F},t)}\  \Fcal \pretp \bound{G}\oplus \bound{B}_1 \iff \Fcal \pretp\bound{G}\oplus \bound{B}_2$.
\end{lemma}



We will need the following simple lemma in the proof of Lemma~\ref{deserves}.

\begin{lemma}
  \label{lem-few-isolated}
  Let ${\cal F}$ be a proper collection, let $t\in \Bbb{N}$, and let ${\bf B} = (B,R,\lambda) \in {\Rcal}^{({\cal F},t)}$.
  Then $B$ contains at most ${\sf size}({\cal F})$ isolated vertices outside of $R$.
\end{lemma}
\begin{proof}
If ${\bf B}={\bf F}^{({\cal F},t)},$ then by definition $B$ is a graph in ${\cal F}$ possibly completed by isolated vertices in $R$, so the lemma follows trivially in this case. So
we may assume that ${\cal F}\not \pretp B$, and suppose
 towards a contradiction, that $B$ contains ${\sf size}({\cal F})+1$ isolated vertices in $V(B) \setminus R$. We define a boundaried graph ${\bf B'}=(B',R',\lambda') \in \Bcal^{(t)}$ such that $R'=R$, $\lambda' = \lambda$, and $B'$ is obtained from $B$ by removing one of the isolated vertices in $V(B) \setminus R$. Note that $|E(B')|=|E(B)|$ and $|V(B')|<|V(B)|$. We claim that ${\bf B}\equiv^{(\Fcal,t)}{\bf B'}$, which contradicts the hypothesis that ${\bf B}  \in {\Rcal}^{({\cal F},t)}$.

 Indeed, consider an arbitrary $\bound{G}\in\Bcal^{(t)}$. Suppose first that $\Fcal \pretp \bound{B'}\oplus \bound{G}$. Since $B'$ is a subgraph of $B$, $\bound{B'}\oplus \bound{G}$ is a subgraph of $\bound{B}\oplus \bound{G}$, hence $\Fcal \pretp \bound{B}\oplus \bound{G}$ as well. Conversely, suppose that $H \pretp \bound{B}\oplus \bound{G}$ for some graph $H \in \Fcal$. Since all the graphs in $\Fcal$, in particular $H$, have at most ${\sf size}(\Fcal)$ vertices, any topological minor model of  $H$ in $\bound{B}\oplus \bound{G}$ uses at most ${\sf size}(\Fcal)$ isolated vertices in $V(B)\setminus R$. Therefore, by possibly using another isolated vertex in $V(B')\setminus R$ instead of the removed one, $\bound{B'}\oplus \bound{G}$ also contains a topological minor model of $H$, hence $\Fcal \pretp \bound{B'}\oplus \bound{G}$.
\end{proof}

\medskip
\noindent
\textbf{Folios.}
Let ${\cal F}$ be a proper collection.
Given $t,r\in\Bbb{N},$ we define ${\cal A}_{{\cal F},r}^{(t)}$ as the set of all
pairwise  non-isomorphic boundaried graphs
that contain at most $r$ non-boundary vertices,  whose
label set is a subset of $\intv{1,t},$ and whose underlying graph belongs to $\ex_{\sf tm}({\cal F})$.
Note that a graph in ${\cal A}_{{\cal F},r}^{(t)}$ is not necessarily a $t$-boundaried graph.

Given a $t$-boundaried graph ${\bf B}$ and an integer $r\in\Bbb{N},$
we define  the \emph{$({\cal F},r)$-folio} of ${\bf B},$ denoted by  ${\sf folio}({\bf B},{\cal F},r),$ as the set
containing all boundaried graphs  in ${\cal A}^{(t)}_{{\cal F},r}$ that are topological minors of ${\bf B}$.
Moreover, in case ${\cal F}\pretp{\bf B},$ we also include in ${\sf folio}({\bf B},{\cal F},r)$
the graph ${\bf F}^{({\cal F},t)}$.

We also define
$\frak{F}_{{\cal F},r}^{(t)}=2^{{\cal A}_{{\cal F},r}^{(t)}\cup\{{\bf F}^{({\cal F},t)}\}}$ and notice that $\{{\sf folio}({\bf B},{\cal F},r)\mid \bound{B}\in {\cal B}^{(t)}\}\subseteq \frak{F}_{{\cal F},r}^{(t)},$ i.e., $\frak{F}_{{\cal F},r}^{(t)}$ contains all  different  $({\cal F},r)$-folios of $t$-boundaried graphs.

\begin{lemma}
  \label{humanism}
  Let $t\in\Nbb$ and let ${\cal F}$ be a proper collection.
  For every  $t$-boundaried graph ${\bf B}$ and every $r\in \Nbb,$  it holds that $|{\sf folio}({\bf B},{\cal F},r)|=2^{\Ocal_{r+d}(t\log t)},$ where $d={\sf size}({\cal F})$.
  Moreover, $|\frak{F}_{{\cal F},r}^{(t)}|=2^{2^{{\Ocal_{r+d}(t\log t)}}}$.
\end{lemma}
\begin{proof}
  Let $t \in \Nbb,$ let $\Fcal$ be a proper collection, let $r \in \Nbb,$ and let $n = t+r$.
  We prove a stronger result, namely that $|{\cal A}^{(t)}_{{\cal F},r}|=2^{\Ocal_{r+d}(t\log t)}$. The claimed bound on $|\frak{F}_{{\cal F},r}^{(t)}|$ then follows directly by definition of the set $\frak{F}_{{\cal F},r}^{(t)}$. By~\cite{KomloS96}, there exists a constant $c$ such that
  for each $G \in \ex_{\sf tm}(\Fcal),$  $|E(G)| \leq c \cdot |V(G)|$.
  By definition, every underlying graph of an element of ${\cal A}^{(t)}_{{\cal F},r}$ is in $\ex_{\sf tm}(\Fcal)$.
  If we want to construct an element $\Bound{G} = (G,R,\lambda)$ of ${\cal A}^{(t)}_{{\cal F},r}$ with at most $n$ vertices, then there are asymptotically at most $c\cdot n \cdot {n^2 \choose c\cdot n} \leq c \cdot n^{1+2\cdot c\cdot n}$ choices for the edge set $E(G),$ at most $t\cdot {n \choose t} \leq t \cdot n^t$ choices for $R,$ and $t^{|R|} \leq t^t$ choices for the function $\lambda$.
  We obtain that ${\cal A}^{(t)}_{{\cal F},r}$ is of size at most $n \cdot 2^{(1+2\cdot c\cdot n) \log n }\cdot 2^{t \log t} =2^{\Ocal_{r+d}(t\log t)} ,$ and
  the lemma follows.
\end{proof}

The following lemma indicates that folios define a refinement of the equivalence relation $\equiv^{({\cal F},t)}$.

\begin{lemma}
  \label{derailed}
  Let $\Fcal$ be a proper collection and let $d={\sf size}(\Fcal)$.
  Let also ${\bf B}_1$ and ${\bf B}_{2}$ be two $t$-boundaried graphs.
  If  ${\sf folio}({\bf B}_{1},{\cal F},d)={\sf folio}({\bf B}_{2},{\cal F},d),$ then ${\bf B}_{1}\equiv^{({\cal F},t)}{\bf B}_{2}$.
\end{lemma}

\begin{proof}
  Let ${\bf B}_1$ and ${\bf B}_{2}$ be two $t$-boundaried graphs such that
  ${\sf folio}({\bf B}_{1},{\cal F},d)={\sf folio}({\bf B}_{2},{\cal F},d)$.
  We fix $\bound{G}\in\Bcal^{(t)},$ and we need to prove that $\Fcal \pretp \bound{G}\oplus \bound{B}_1$  if and only if $\Fcal \pretp\bound{G}\oplus \bound{B}_2$.

  Assume first that $\Fcal \pretp \bound{G}\oplus \bound{B}_1$. Then there exists a graph $F \in \Fcal$ and a  topological minor model  $(\phi,\sigma)$ of $F$ in $\bound{G}\oplus \bound{B}_1$. This topological minor model  $(\phi,\sigma)$ can be naturally decomposed into two topological minor models $(\phi_0,\sigma_0)$ and $(\phi_1,\sigma_1)$ of two graphs $F_0$ and $F_1$ in ${\cal A}_{{\cal F},d}^{(t)},$ respectively, with $F_0 \odot F_1 = F,$ such that $(\phi_0,\sigma_0)$ (resp. $(\phi_1,\sigma_1)$) is a topological minor model of $F_0$ (resp. $F_1$) in the (boundaried) graph $\bound{G}$ (resp. $\bound{B}_1$). Since ${\sf folio}({\bf B}_{1},{\cal F},d)={\sf folio}({\bf B}_{2},{\cal F},d),$ there exists a topological minor model $(\phi_2,\sigma_2)$ of $F_1$ in $\bound{B}_2$. Combining the topological minor models $(\phi_0,\sigma_0)$ and $(\phi_2,\sigma_2)$ gives rise to a topological minor model $(\phi',\sigma')$ of $F$ in $\bound{G}\oplus \bound{B}_2,$ and therefore $\Fcal \pretp \bound{G}\oplus \bound{B}_2$.

  Conversely, assume that $\Fcal \not \pretp \bound{G}\oplus \bound{B}_1,$ and assume for contradiction that there exists a graph $F \in \Fcal$ and a topological minor model  $(\phi,\sigma)$ of $F$ in $\bound{G}\oplus \bound{B}_2$. Using the same arguments as above, $(\phi,\sigma)$ implies the existence of a  topological minor model  $(\phi',\sigma')$ of $F$ in $\bound{G}\oplus \bound{B}_1,$ contradicting the hypothesis that $\Fcal \not \pretp \bound{G}\oplus \bound{B}_1$.
\end{proof}

Lemmas~\ref{humanism} and~\ref{derailed} directly imply the following.

\begin{lemma}
  \label{entering}
  There exists a function $h_{1}:\Bbb{N}\times\Bbb{N}\to\Bbb{N}$ such that
  if ${\cal F}$ is a proper collection
  and $t\in\Bbb{N},$ then $|{\Rcal}^{( {\cal F},t)}|\leq h_{1}(d,t)$ where $d={\sf size}({\cal F})$. Moreover $h_{1}(d,t)=2^{2^{\Ocal_d(t\cdot \log t)}}$.
\end{lemma}

\section{Branch decompositions of boundaried graphs}
\label{dictates}

Let $\bound{G}=(G,R,\lambda)$ be a boundaried graph
and let  $\rho$ be a vertex labeling of $G$
where $\lambda\subseteq \rho$. A {\em  branch decomposition} of  $\bound{G}$ is a pair  $(T, \sigma)$ where  $T$ is a ternary tree  and
$\sigma : E(G)\cup\{R\} \rightarrow L(T)$ is a bijection. Let $r=\sigma(R)$ and let  $e_{r}$ be the {unique} edge in $T$ that is incident to $r$. We call $r$ {\em the root}
of $T$.
Given an edge $e\in E(T),$
we define $T_{e}$  as the one of the two connected components of $T \bs \{e\}$
that does not contain the root $r$. We then define ${\bf G}_{e}=(G_{e},R_{e},\lambda_{e})$
where $E(G_{e})=\sigma^{-1}(L(T_{e})\cap L(T)),$  $V(G_{e})=\cupall E(G_{e}),$
$R_{e}$ is the set containing every vertex of $G$ that is an endpoint
of an edge in $E(G_{e})$
and also belongs to a set in $\{R\}\cup(E(G)\setminus E(G_{e}))$ (here we treat edges in $E(G)\setminus E(G_{e})$ as 2-element sets), and $\
\lambda_{e}=\rho|_{R_{e}}$, i.e., $\rho$ serves as a universal labeling of $G$ that
imposes a labeling of the vertices of all boundaried graphs that are obtained from subgraphs of $G$. We also set $t_{e}=|R_{e}|$ and observe that ${\bf G}_{e}$
is a $t_{e}$-boundaried graph.
The \emph{width} of $(T,\sigma)$ is $\max\{t_{e}\mid e\in E(T)\}$.
The \emph{branchwidth} of ${\bf G},$ denoted by $\bw({\bf G}),$ is the minimum width over all branch decompositions of ${\bf G}$.

This is an  extension of the definition of a branch
decomposition on graphs, given in~\cite{RobertsonS91}, to boundaried graphs.
%
%
Indeed, if $G$ is a graph, then a {\em branch decomposition} of $G$ is
a branch decomposition of $(G,\emptyset,\varnothing)$. We also
define the {\em branchwidth} of $G$ as $\bw(G)=\bw(G,\emptyset,\varnothing)$.

\begin{lemma}
  \label{likewise}
  Let  $\bound{G}=(G,R,\lambda)$ be a boundaried graph.
  Then $\bw(\bound{G})\leq \bw(G)+|R|$.
\end{lemma}

\begin{proof}
  Let $(T',\sigma')$ be a branch decomposition of ${\bf G}'=(G,\emptyset,\varnothing)$
  and let $r$ be the root of $T'$. Recall that ${\bf G}_{e}'=(G_{e}',R_{e}',\lambda'_{e}), e\in E(T')$.
  We construct a branch decomposition
  $(T,\sigma)$ of ${\bf G}=(G,R,\lambda)$ as follows: we set $T=T'$
  and $\sigma=(\sigma'\setminus \{(\emptyset,r)\})\cup\{(R,r)\}$.
  Note that ${\bf G}_{e}=(G_{e}',R_{e},\lambda_{e}), e\in E(T),$ where
  $R_{e}\subseteq R_{e}'\cup R$. This means that $|R_{e}|\leq  |R_{e}'|+|R|,$ therefore
  $\bw(\bound{G})\leq \bw(G)+|R|$.
\end{proof}



The following lemma is a combination of the single-exponential linear-time constant-factor approximation of treewidth by Bodlaender et al.~\cite{BodlaenderDDFLP16},  with the fact that any graph $G$ with $|E(G)| \geq 3$ satisfies that $\bw(G) \leq \tw(G) + 1 \leq \frac{3}{2}\bw(G)$~\cite{RobertsonS91}; it is worth noting that from the proofs of these inequalities, simple polynomial-time algorithms for transforming a branch (resp. tree) decomposition into a  tree (resp. branch) decomposition can be derived.

\begin{lemma}
  \label{initials}
  There exists an algorithm that receives as input  a graph $G$ and a $w\in\Bbb{N}$
  and  either reports that $\bw(G)> w$ or outputs a branch decomposition  $(T,\sigma)$
  of $G$ of width $\Ocal(w)$.
  Moreover, this algorithm runs in $2^{\Ocal(w)}\cdot n$ steps.
\end{lemma}

\begin{lemma}
  \label{granting}
  There exists a function $\mu: \Bbb{N}\to\Bbb{N}$ such that
  for every planar subcubic collection  ${\cal F},$
  every graph in ${\sf ex}_{\sf tm}({\cal F})$  has  branchwidth at most  $y=\Chu(d)$ where $d={\sf size}({\cal F})$.
\end{lemma}

\begin{proof}
  Let $G \in {\sf ex}_{\sf tm}({\cal F})$ and let $F \in {\cal F}$ be a planar subcubic graph. Since $F$ is subcubic and $F \npreceq_{\sf tm} G,$ it follows (see~\cite{Die10}) that $F \npreceq_{\sf m} G,$ and since $F$ is planar this implies by~\cite{RobertsonS86} that $\tw(G),$ hence $\bw(G)$ as well, is bounded by a function depending only on $F$.
\end{proof}

\section{Proof of Theorem~\ref{perishes}}
\label{redeemed}

We already have all the ingredients to prove Theorem~\ref{perishes}.

\begin{proof}[Proof of Theorem~\ref{perishes}]
  We provide a dynamic programming algorithm
  for the computation of ${\bf tm}_{\cal F}(G)$
  for the general case where ${\cal F}$ is a proper collection.
  We first
  consider an, arbitrarily chosen, vertex labeling $
  \rho$ of $G$.
  From Lemma~\ref{initials}, we may
  assume that we have a   branch decomposition $(T,\sigma)$ of $(G,\emptyset,\varnothing)$ of
  width $\Ocal(w),$ where $w=\tw(G)$.
  This gives rise to the $t_{e}$-boundaried
  graphs  ${\bf G}_{e}=(G_{e},R_{e},\lambda_{e})$ for each $e\in E(T)$. Moreover,
  if $r$ is the root of $T,$ $\sigma^{-1}(r)=\emptyset=R_{e_{r}}$
  and  $\bound{G}_{e_{r}}=(G,\emptyset,\varnothing)$.  Keep also in mind that $t_{e}=\Ocal(\tw(G))$ for every $e\in E(T)$.

  For each $e\in E(T),$ we say that $(L,{\cal C})$ is an {\em $e$-pair}
  if $L\subseteq R_{e}$ and ${\cal C}\in \frak{F}_{{\cal F}
    ,d}^{(t'_e)}$ where $t'_e=t_e-|L|$. We also denote by ${\cal P}_{e}$ the set of all $e$-pairs.
  Clearly, $|{\cal P}_{e}|=\sum_{i\in[0,t_{e}]}\binom{t_e}{i}\cdot |\frak{F}_{{\cal F},d}^{(t_{e}-i)}|,$ and therefore, from Lemma~\ref{humanism}, $|{\cal P}_{e}|=2^{2^{{\Ocal_{d}(w\log w)}}}$.

  We then define the function ${\bf tm}^{(e)}_{\cal F}: {\cal P}_{e}\to\Bbb{N} \cup\{\infty\}$
  such that if $(L,{\cal C})\in {\cal P}_{e},$ then
  $${\bf tm}^{(e)}_{\cal F}(L,{\cal C})=\min\{|S|\mid S \subseteq V({G}_{e})~ \wedge ~L=R_{e}\cap S~ \wedge ~{\cal C}={\sf folio}({\bf G}_{e}\setminus S,d)\}.$$
  In the above definition, if such a set $S$ does not exist, we set the value of the function to $\infty$. Note that ${\cal P}_{e_{r}}=\{\emptyset\}  \times \frak{F}_{{\cal F}
    ,d}^{(0)}$.
  Note also that the set ${\cal A}_{{\cal F},d}^{(0)}$
  contains only those graphs that do {\sl not} contain some graph in ${\cal F}$ as a topological minor. Therefore
  $${\bf tm}_{\cal F}(G)=\min\{  {\bf tm}^{(e_r)}_{\cal F}(\emptyset,{\cal C})  \mid {\cal C}\in 2^{{\cal A}_{{\cal F},d}^{(0)}}\}.$$

  Hence, our aim is to give a way to compute ${\bf tm}^{(e)}_{\cal F}$ for every $e\in E(T)$.
  Our dynamic programming algorithm does this in a bottom-up fashion, starting from the edges that contain as endpoints leaves of $T$ that are different from the root.
  Let $\ell\in L(T)\setminus \{r\}$ and
  let $e_{\ell}$ be the {unique} edge of $T$ that contains it.
  Let also $\sigma^{-1}(\ell)=\{x,y\}$.
  Clearly, ${\bf G}_{e_{\ell}}=(\{x,y\},\{\{x,y\}\})$ and

  $${\cal P}_{e_{\ell}}=\big\{(\{x,y\} \times \frak{F}_{{\cal F}
    ,d}^{(0)})\big\}\cup(\big\{\{x\},\{y\}\big\}\times \frak{F}_{{\cal F},d}^{(1)})\cup (\{\emptyset\}\times\frak{F}_{{\cal F},d}^{(2)}).$$
  As the size of the elements in ${\cal P}_{e_{\ell}}$ depends only on  $d,$ it is possible to compute ${\bf tm}^{(e_{\ell})}_{\cal F}$ in $\Ocal_{d}(1)$ steps.

  Let $e\in \{e_{r}\}\cup E(T\setminus L(T)),$ and let $e_{1}$ and $e_{2}$ be
  the two other edges of $T$ that share an endpoint with $e$ and where each path from them to $r$ contains $e$. We also set $$F_{e}=\big(R_{e_{1}}\cup R_{e_{2}}\big)\setminus R_{e}.$$
  For the dynamic programming algorithm, it is enough to describe how to compute ${\bf tm}_{\cal F}^{(e)}$ given ${\bf tm}_{\cal F}^{(e_i)}, i\in\intv{1,2}$.
  For this, given an $e$-pair $(L,{\cal C})\in{\cal P}_{e}$ it is possible to verify  that
  \begin{eqnarray*}
    {\bf tm}^{(e)}_{\cal F}(L,{\cal C}) & = & \min\big\{{\bf tm}_{\cal F}^{(e_{1})}(L_1,{\cal C}_{1})+{\bf tm}_{\cal F}^{(e_{2})}(L_2,{\cal C}_{2})-|L_{1}\cap L_{2}|\mid \\
                                        & & ~~~~~~ \mbox{$(L_{i},{\cal C}_{i})\in{\cal P}_{e_{i}},i\in\intv{1,2},$} \\
                                        & & ~~~~~~ L_{i}\setminus F_{e}=L\cap R_{e_{i}}, i\in\intv{1,2},\\
                                        & & ~~~~~~ L_{1}\cap R_{e_1} \cap R_{e_2}=L_{2}\cap R_{e_1} \cap R_{e_2} , \mbox{~and~} \\
                                        & & ~~~~~~ {\cal C}=\!\!\!\bigcup_{({\bf B}_{1},{\bf B}_{2})\in {\cal C}_{1}\times{\cal C}_{2}}\!\!\!{\sf folio}\Big(\big(({\bf B}_{1}\diamond Z_{1})\odot({\bf B}_{2}\diamond  Z_{2})\big)|_{Z},{\cal F}, t_e - |L|\Big)  \\
                                        & & ~~~~~~~~~~\mbox{\ where\ } \mbox{$Z=\rho(R_{e}\setminus L)$ and $Z_{i}=\rho(R_{e_{i}}\setminus L_{i}), i\in\intv{1,2}$}
                                            \big\}.
  \end{eqnarray*}

  Note that given ${\bf tm}^{(e_i)}_{\cal F}, i\in\intv{1,2}$ and a $(L,{\bf B})\in{\cal P}_{e},$
  the value of ${\bf tm}^{(e)}_{\cal F}(L,{\bf B})$ can be computed by the above formula in  $\Ocal_d(|{\cal P}_{e_1}|\cdot |{\cal P}_{e_2}|)=2^{2^{{\Ocal_{d}(w\log w)}}}$ steps.
  As $|{\cal P}_{e}|=2^{2^{{\Ocal_{d}(w\log w)}}},$ the computation of the function ${\bf tm}^{(e)}_{\cal F}$ requires again $2^{2^{{\Ocal_{d}(w\log w)}}}$ steps. This means that the whole dynamic programming requires
  $2^{2^{\Ocal_{d}(w\cdot \log w)}}\cdot |V(T)| = 2^{2^{{\Ocal_{d}(w\log w)}}}\cdot  |E(G)|$ steps. As $|E(G)|=\Ocal(\tw(G)\cdot |V(G)|),$ the claimed running time follows.
\end{proof}

%

\section{Improved bounds when excluding a planar graph}
\label{expected}
We now prove the following result.

\begin{lemma}
  \label{peepshow}
  Let $t\in\Bbb{N}$ and ${\cal F}$ be a proper collection containing a subcubic planar graph, where $d={\sf size}({\cal F}),$
  and let ${\Rcal}^{({\cal F},t)}$ be a set of representatives for $\equiv^{(\Fcal,t)}$.
  Then $|{\Rcal}^{({\cal F},t)}|=2^{\Ocal_{d}(t\cdot \log t)}$.
  Moreover, there exists an algorithm that given ${\cal F}$ and $t,$
  constructs a set of representatives ${\Rcal}^{({\cal F},t)}$ in $2^{\Ocal_{d}(t\cdot \log t)}$ steps.
\end{lemma}

Before we proceed with the proof of Lemma~\ref{peepshow},  we need
a series of results. The proof of the following lemma uses ideas similar to the ones presented by Garnero et al.~\cite{GarneroPST15}.


%

%
%
%

%

\begin{lemma}
  \label{deserves}
  There is a  function $h_2:\Bbb{N}\times\Bbb{N}\to\Bbb{N}$
  such that if ${\cal F}$ is a proper collection containing a subcubic planar graph,  where
  $d={\sf size}({\cal F}),$
  $t\in\Bbb{N},$
  ${\bf B}=(B,R,\lambda)\in{\Rcal}^{({\cal F},t)}\setminus \{{\bf F}^{({\cal F},t)}\},$
  $z\in \mathbb{N},$ and $X$ is a subset of $V(B)$ such that  $X\cap R=\emptyset$ and $|N_{B}(X)|\leq z,$ then $|X|\leq h_{2}(z,d)$.
\end{lemma}

\begin{proof}

  We set $h_{2}(z,d)=2^{h_{1}(d,\Chu(d)+z)\cdot (z+\mu(d)+1)+\zeta(\Chu(d)+z)}+z+d,$ where $h_{1}$ is the function of Lemma~\ref{entering}, $\mu$ is the function of Lemma~\ref{granting}, and $\zeta:\Bbb{N}\to\Bbb{N}$ is defined as $\zeta(x)=2^{x \choose 2}$.
  Let  $y=\Chu(d),$  $q=h_{1}(d,y+z)\cdot (x+y+1)\cdot \zeta(y+z),$ $s=h_{2}(z,d),$ and observe that $s=2^{q}+z+d$.
  Towards a contradiction, we assume that $|X|>s$.

  \smallskip
  Let ${\bf B}=(B,R,\lambda)\in{\Rcal}^{({\cal F},t)}\setminus \{{\bf F}^{({\cal F},t)}\}$ and let $\rho$ be a vertex-labeling of $B$ where $\lambda\subseteq \rho$.
  As ${\bf B}\neq {\bf F}^{({\cal F},t)},$ it follows that
  \begin{eqnarray}
    & B \in  {\sf ex}_{\sf tm}({\cal F}). & \label{indicate}
  \end{eqnarray}
  %
  We set $G=B[X\cup N_{B}(X)]$ and observe that $|V(G)|\geq |X|>s$.
  As $G$ is a subgraph of $B,$ \eqref{indicate} implies  that
  \begin{eqnarray}
    & G \in  {\sf ex}_{\sf tm}({\cal F}), &
  \end{eqnarray}
  and therefore, from Lemma~\ref{granting}, $\bw(G)\leq y$.
  Let $R'=N_{B}(X)$ and $\lambda'=\rho|_{R'}$.
  We set ${\bf G}=(G,R',\lambda')$. From Lemma~\ref{likewise},
  $\bw({\bf G})\leq \bw(G)+|R'|\leq y+|R'|=y+z$.
  \smallskip

  We now proceed to bound the number of isolated vertices of $G$. Note that every isolated vertex of $G$ either belongs to $R'$ or belongs to $X$ and  was already an isolated vertex in $B$. Since, by Lemma~\ref{lem-few-isolated}, $B$ has at most $d$ isolated vertices in $V(B) \setminus R$, and $X \cap B = \emptyset$, it follows that $G$ has at most $|R'| + d$ isolated vertices.

  Hence, since for any connected component $C$ of $G$ that is not an isolated vertex it holds that $|E(C)| \geq |V(C)|/2$, we conclude that

%
   \begin{eqnarray}\label{eq-number-edges}
    |E(G)|\ \geq\ \frac{|V(G)|-|R'|-d}{2} \ \geq \ \frac{|V(G)|-z-d}{2} \ > \ \frac{s-z-d}{2} \ =\ 2^{q-1}.
  \end{eqnarray}

  Let $(T,\sigma)$ be a branch decomposition of ${\bf G}$
  of width at most $y+z$. We also consider the graph ${\bf G}_{e}=(G_{e},R_e,\lambda_e),$ for each $e\in E(T)$ (recall that $\lambda_{e}\subseteq \rho$).
  Observe  that
  \begin{eqnarray}
    \forall e\in E(T),\ |R_{e}|\leq y+z.\label{withhold}
  \end{eqnarray}
  We define  ${\cal H}=\{{\sf rep}_{\cal F}({\bf G}_{e})\mid e\in E(T)\}$.
  From~\eqref{withhold},
  ${\cal H}\subseteq \bigcup_{i\in\intv{0,y+z}}{\Rcal}^{({\cal F},i)}$.
  From Lemma~\ref{entering}, $|{\cal H}|\leq  (y+z+1)\cdot h_{1}(d,y+z),$ therefore $q\geq |{\cal H}|\cdot\zeta(y+z)$.
  Let $r$ be the root of $T$ and let $P$ be a longest path in $T$ that has $r$ as an endpoint.
  As by (\ref{eq-number-edges}), $G$ has more than  $2^{q-1}$ edges, $T$ also has more than $2^{q-1}$ leaves different from $r$.
  This means that  $P$ has more than $q$ edges. Recall that $q\geq |{\cal H}|\cdot\zeta(y+z)$.
  As a consequence, there
  is a set ${\cal S}\subseteq \{{\bf G}_{e}\mid e\in E(P)\}$
  where $|{\cal S}|>\zeta(y+z)$ and ${\sf rep}_{\cal F}({\cal S})$ contains
  only one boundaried graph (i.e., all the
  boundaried graphs in ${\cal S}$ have the same ${\cal F}$-representative).
  From Observation~\ref{elegance},
  there are two graphs ${\bf G}_{e_{1}},{\bf G}_{e_{2}}\in {\cal S},$ $e_{1}\neq e_{2},$ such that
  \begin{eqnarray}
    {\bf G}_{e_1}&\!\!\! \equiv^{({\cal F},t)}&\!\!\! {\bf G}_{e_2}\text{~and} \label{workshop}\\
    {\bf G}_{e_1} &\!\!\! \sim&\!\!\! {\bf G}_{e_2}. \label{thinkers}
  \end{eqnarray}
  W.l.o.g., we assume that $e_{1}$ is in the path in $T$ between $r$ and some endpoint of $e_{2}$. This implies that  the underlying graph of
  ${\bf G}_{e_{1}}$ is a proper subgraph of the underlying graph of
  ${\bf G}_{e_{2}},$ therefore
  %
  \begin{eqnarray}
    & |E({\bf G}_{e_{2}})|<|E({\bf G}_{e_{1}})|.& \label{deciding}
  \end{eqnarray}
  Recall that  ${\bf G}_{e_{i}}=(G_{e_i},R_{e_i},\lambda_{e_{i}}), i\in\intv{1,2}$.
  Let $B^-=B\setminus (V(G_{e_{1}})\setminus R_{e_{1}})$
  and we set ${\bf B}^{-}=(B^-,R_{e_{1}},\lambda_{e_{1}})$.
  Clearly, $ {\bf B}^-\sim{\bf G}_{e_{1}}$. This, combined with~\eqref{thinkers},  implies
  that
  \begin{eqnarray}
    & {\bf B}^-\sim{\bf G}_{e_{2}}.
    & \label{mccarthy}
  \end{eqnarray}
  Let now $B^*={\bf B}^-\oplus {\bf G}_{e_{2}}$.
  Combining~\eqref{deciding} and~\eqref{mccarthy}, we may deduce that
  \begin{eqnarray}
    |E(B^*)|<|E(B)|. 
    \label{merciful}
  \end{eqnarray}
  We now set ${\bf B}^{*}=(B^*,R,\lambda)$
  and recall that $t=|R|$. Clearly, both ${\bf B}$ and ${\bf B}^{*}$
  belong to ${\cal B}^{(t)}$.
  \smallskip

  We now claim that  ${\bf B}\equiv^{({\cal F},t)}{\bf B}^{*}$.
  For this, we consider any ${\bf D}=(D,R,\lambda)\in{\cal B}^{(t)}$. We
  define ${\bf B}^{\star}=(B^-,R,\lambda),$
  $D^+={\bf D}\oplus {\bf B}^{\star},$ and ${\bf D}^+=(D^+,R_{e_{1}},\lambda_{e_{1}})$.
  Note that
  \begin{eqnarray}
    & {\bf D}\oplus {\bf B}  =  {\bf D}^+\oplus {\bf G}_{e_{1}}  & \text{ and}\label{provoked}\\
    & {\bf D}\oplus {\bf B}^* =  {\bf D}^+\oplus {\bf G}_{e_{2}}.& \label{fruitful}
  \end{eqnarray}
  From~\eqref{workshop},
  we have that ${\cal F}\pretp {\bf D}^+\oplus {\bf G}_{e_{1}}\iff {\cal F}\pretp {\bf D}^+\oplus {\bf G}_{e_{2}}$. This, together with~\eqref{provoked} and~\eqref{fruitful}, implies that ${\cal F}\pretp {\bf D}\oplus {\bf B} \iff {\cal F}\pretp {\bf D}\oplus {\bf B}^*,$ therefore ${\bf B}\equiv^{({\cal F},t)}{\bf B}^{*},$ and the claim follows.
  \smallskip

  We just proved that  ${\bf B}\equiv^{({\cal F},t)}{\bf B}^{*}$. This  together with~\eqref{merciful} contradict the fact that
  ${\bf B}\in{\Rcal}^{({\cal F},t)}$. Therefore $|X|\leq s,$ as required.
\end{proof}
\medskip

Given a graph $G$ and an integer $y,$ we say that a vertex set $S\subseteq V(G)$
is a {\em branchwidth}-$y$-{\em modulator} if $\bw(G\setminus S)\leq y$. This notion is inspired from \emph{treewidth-modulators}, which have been recently used in a series of papers (cf., for instance,~\cite{BodlaenderFLPST16,GarneroPST15,F.V.Fomin:2010oq,KimLPRRSS16line}).


The following proposition is a (weaker) restatement of~\cite[Lemma~3.10 of the full version]{F.V.Fomin:2010oq} (see also~\cite{KimLPRRSS16line}).

\begin{proposition}
  \label{forwards}
  There exists a function $f_{2}:\Bbb{N}_{\geq 1}\times\Bbb{N}\to\Bbb{N}$ such that if  $d\in\Bbb{N}_{\geq 1},$ $y\in\Bbb{N},$
  and $G$ is a graph such that  $G\in{\sf ex}_{\sf tm}(K_{d})$ and
  $G$ contains a {\em branchwidth}-$y$-{\em modulator} $R,$
  then  there exists a partition $\Xcal$ of
  $V(G)$ and an element $X_0 \in \Xcal$ such that $R\subseteq X_0,$ $\max\{|X_0|,|\Xcal|-1\}\leq 2\cdot |R|,$ and for every $X \in \Xcal \sm \{X_0\},$ $|N_{G}(X)|\leq f_{2}(d,y)$.
\end{proposition}

\begin{lemma}
  \label{adorning}
  There is a  function $h_3:\Bbb{N}\to\Bbb{N}$
  such that if $t\in\Bbb{N}$ and ${\cal F}$ is a proper collection containing a subcubic planar graph, where $d={\sf size}({\cal F}),$ then every graph in ${\Rcal}^{({\cal F},t)}$ has at most $t\cdot h_{3}(d)$ vertices.
\end{lemma}

\begin{proof}
  We define $h_3:\Bbb{N}\to\Bbb{N}$ so that $h_{3}(d)=2+h_{2}(f_{2}(d,\Chu(d)),\Chu(d))$
  where $h_{2}$ is the function of Lemma~\ref{deserves}, $f_{2}$ is the function of Proposition~\ref{forwards}, and $\Chu$ is the function of Lemma~\ref{granting}.

  As ${\bf F}^{({\cal F},t)}$ has at most $d$ vertices, we may assume that $\Bound{G} = (G,R, \lambda)\in {\Rcal}^{({\cal F},t)}\setminus \{{\bf F}^{({\cal F},t)}\}$. Note  that  $G \in  {\sf ex}_{\sf tm}({\cal F}),$ therefore, from Lemma~\ref{granting},  $\bw(G)\leq \Chu(d)$. We set $y=\Chu(d)$ and 
  we observe that  $R$ is a {\em branchwidth}-$y$-{\em modulator} of $G$. Therefore, we
  can apply Proposition~\ref{forwards} on $G$ and $R$ and
  obtain a partition
  $\Xcal$ of
  $V(G)$  and an element $X_0 \in \Xcal$ such that
  \begin{eqnarray}
    && R\subseteq X_0,   \label{profanum}\\
    &&  \max\{|X_0|,a\}\leq 2\cdot |R|,  \label{marxists} \text{~and}\\
    && {\forall X \in \Xcal \sm \{X_0\}:\ |N_{G}(X)|\leq f_{2}(d,y). \label{normally}}
  \end{eqnarray}
  From~\eqref{profanum} and~\eqref{normally},
  {each $X \in \Xcal \sm \{X_0\}$}
  is a subset of $V(G)$ such that
  {$X \cap R = \es$}
  and
  { $|N_{G}(X)|\leq f_{2}(d,y)$.}
  Therefore, from Lemma~\ref{deserves},
  {for each $X \in \Xcal\sm \{X_0\},$ $|X|\leq h_{2}(f_{2}(d,y),d)$.}
  We obtain that
  \begin{eqnarray*}
    |G|& = & {|X_0|+\sum_{X \in \Xcal \sm \{X_0\}}|X| }\\
          & \leq^{\eqref{marxists}} &  2\cdot |R| + |R|\cdot h_{2}(f_{2}(d,y),d)\\
          & = & t\cdot (2+h_{2}(f_{2}(d,y),d))\\
          & = & t\cdot h_{3}(d),
  \end{eqnarray*}
  as required.
\end{proof}

The next proposition follows from the  results of Baste et al.~\cite{BNS16} on the number of labeled graphs of bounded treewidth.

\begin{proposition}[Baste et al.~\cite{BNS16}]
  \label{touching}
  Let $n,q\in\Bbb{N}$. The number of labeled graphs with at most $n$ vertices and branchwidth at most $q$ is $2^{\Ocal_{q}(n\cdot \log n)}$.
\end{proposition}

We are now ready to prove Lemma~\ref{peepshow}.

\begin{proof}[Proof of Lemma~\ref{peepshow}]

  Before we proceed to the proof we need one more definition. Given $n\in\Bbb{N},$ we set  ${\cal B}^{({\cal F},t)}_{\leq n}={\cal A}_{{\cal F},n-t}^{(t)}\cup \{{\bf F}^{({\cal F},t)}\}$.

  %

  Note that, from Lemma~\ref{adorning},
  ${\Rcal}^{({\cal F},t)}\subseteq {\cal B}^{({\cal F},t)}_{\leq n},$
  where  $n =  t\cdot h_{3}(d)$. Also, from Lemma~\ref{granting}, all graphs in ${\cal B}^{({\cal F},t)}_{\leq n}$ have branchwidth at most $y= \max\{\Chu(d),t\}$.
  The   fact that $|{\cal B}^{({\cal F},t)}_{\leq n}|= 2^{\Ocal_{d}(t\cdot \log t)}$ follows easily by applying Proposition~\ref{touching}
  for  $n$ and $q$.

  The algorithm claimed in the second statement of the lemma constructs a set of representatives ${\Rcal}^{({\cal F},t)}$ as follows: first it finds a partition ${\cal Q}$
  of  ${\cal B}^{({\cal F},t)}_{\leq n}$ into
  equivalence classes with respect to  $\equiv^{(\Fcal,t)}$ and then picks  an element with minimum number of edges from each set of this partition.

  The computation of the above partition of  ${\cal B}^{({\cal F},t)}_{\leq n}$
  is based on the fact that,
  given two $t$-boundaried graphs ${\bf B}_1$ and ${\bf B}_{2},$
  ${\bf B}_1\equiv^{(\Fcal,t)}{\bf B}_{2}$ if and only if, for every ${\bf G}\in {\cal B}^{({\cal F},t)}_{\leq n}$, ${\cal F}\pretp{\bf G}\oplus {\bf B}_{1}\iff {\cal F}\pretp{\bf G}\oplus {\bf B}_{2}$. This fact follows directly from  Lemma~\ref{proceeds} and taking into account that ${\Rcal}^{({\cal F},t)}\subseteq {\cal B}^{({\cal F},t)}_{\leq n}$.

  Note that it takes $|{\cal B}^{({\cal F},t)}_{\leq n}|^{3}\cdot \Ocal_{d}(1)\cdot t^{\Ocal(1)}$ steps
  to construct ${\cal Q}$, by using the topological minor containment algorithm of Grohe et al.~\cite{GroheKMW11}.
  As $|{\cal B}^{({\cal F},t)}_{\leq n}|= 2^{\Ocal_{d}(t\cdot \log t)},$ the construction  of ${\cal Q},$ and therefore of ${\Rcal}^{({\cal F},t)}$ as well,
  can be done in the claimed number of steps.
\end{proof}

\section{Proof of Theorem~\ref{oblivion}}
\label{dramatic}


We are now ready to prove Theorem~\ref{oblivion}. The main difference with respect to the proof of Theorem~\ref{perishes} is an improvement on the size of the tables of the dynamic programming algorithm, namely $|{\cal P}_{e}|,$ where the fact that the collection $\Fcal$ contains a  planar subcubic graph is exploited in order to bound the treewidth.

\begin{proof}[Proof of Theorem~\ref{oblivion}]
  We provide a dynamic programming algorithm
  for the computation of ${\bf tm}_{\cal F}(G)$.   We first
  consider an, arbitrarily chosen, vertex labeling $
  \rho$ of $G$.
  From Lemma~\ref{initials}, we may
  assume that we have a   branch decomposition $(T,\sigma)$ of $(G,\emptyset,\varnothing)$ of
  width at most $w=\Ocal(\bw(G)) = \Ocal(\tw(G))$.
  This gives rise to the $t_{e}$-boundaried
  graphs  ${\bf G}_{e}=(G_{e},R_{e},\lambda_{e})$ for each $e\in E(T)$. Moreover,
  if $r$ is the root of $T,$ $\sigma^{-1}(r)=\emptyset=R_{e_{r}}$
  and  $\bound{G}_{e_{r}}=(G,\emptyset,\varnothing)$. Keep also in mind that $t_{e}=\Ocal(\tw(G))$ for every $e\in E(T)$.


  Our next step is to define the tables of the dynamic programming algorithm. For a positive integer $t$, we let $\overline{\cal R}^{({\cal F},t)} = {\cal R}^{({\cal F},t)}\setminus\{{\bf F}^{({\cal F},t)}\}$.
  Let $e\in E(T)$. We call the pair $(L,{\bf B})$ an \emph{$e$-pair}
  if
  \begin{enumerate}
  \item $L\subseteq R_{e}$, and
  \item ${\bf B}=(B,R,\lambda)\in\overline{\cal R}^{({\cal F},k')}$  where $k'=|R_{e}\setminus L|=t_{e}-|L|$.
  \end{enumerate}
  For each $e\in E(T),$ we denote by ${\cal P}_{e}$  the set of all $e$-pairs.   Note that
  \begin{eqnarray*}
    |{\cal P}_{e}| & = & \sum_{i\in \intv{0,t_{e}}}\binom{t_e}{i}\cdot |\overline{\cal R}^{({\cal F},t_{e}-i)}|\\
                   & = &(t_{e}+1) \cdot 2^{t_{e}}\cdot 2^{\Ocal_{d}(t_{e}\cdot \log t_{e})} \text{~~~~(from Lemma~\ref{peepshow})}
    \\
                   & = &2^{\Ocal_{d}(w\cdot \log w)}.
  \end{eqnarray*}

  We then define the function ${\bf tm}^{(e)}_{\cal F}: {\cal P}_{e}\to\Bbb{N} \cup\{\infty\}$ such that if $(L,{\bf B})\in {\cal P}_{e},$ then
  $${\bf tm}^{(e)}_{\cal F}(L,{\bf B})=\min\{|S|\mid S \subseteq V({G}_{e}) \wedge L=R_{e}\cap S~ \wedge ~{\bf B}={\sf rep}_{\cal F}({\bf G}_{e}\setminus S)\},$$ %


where the value `$\infty$' is assigned when such a set $S$ does not exist.

\noindent Note that  ${\cal P}_{e_{r}} = \{(\emptyset,{\bf B}) \mid {\bf B}  \in \overline{\cal R}^{({\cal F},0)}\}$, and therefore

\begin{eqnarray*}
    {\bf tm}_{\cal F}(G) & = &   \min\{|S|\mid S  \subseteq V(G)\ \wedge \ {\cal F}\not \pretp G\setminus S\}~~~~~ ~~ ~~ ~~ ~~ ~~ ~~ ~~ ~~ ~~ ~~ ~~ ~~ ~ \text{(from Equation~\eqref{overlaid})}\\
                         & = &  \min\{|S|\mid S  \subseteq V(G_{e_r})\ \wedge \ \emptyset = R_{e_r} \cap S\ \wedge\ {\cal F}\not \pretp G_{e_r}\setminus S \}\\
                        & = & \min_{{\bf B} \in  \overline{\cal R}^{({\cal F},0)}} \{       \min\{|S|\mid S  \subseteq V(G_{e_r})\ \wedge \ \emptyset = R_{e_r} \cap S\ \wedge\         {\bf B}= {\sf rep}_{\cal F}((G_{e_r}\setminus S,\emptyset,\varnothing))    \}\\
                         & = & \min_{{\bf B} \in  \overline{\cal R}^{({\cal F},0)}} \{    {\bf tm}^{(e_{r})}_{\cal F}(\emptyset,{\bf B})    \}.
 \end{eqnarray*}



  Therefore, in order to compute $ {\bf tm}_{\cal F}(G)$, it is enough to compute ${\bf tm}^{(e)}_{\cal F}$ for every $e\in E(T)$. Note that, by Lemma~\ref{peepshow}, we may assume that we have at hand the set ${\Rcal}^{({\cal F},t)}$ of representatives for every $t \leq w$.
  Our dynamic programming algorithm does this in a bottom-up fashion, starting from the edges that contain as endpoints leaves of $T$ that are different to the root.
  Let $l\in L(T)\setminus \{r\}$ and
  let $e_{\ell}$ be the edge of $T$ that contains it.
  Let also $\sigma^{-1}(\ell)=\{x,y\}$.
  Clearly, ${\bf G}_{e_{\ell}}=(\{x,y\},\{\{x,y\}\})$ and
  $${\cal P}_{e_{\ell}}=\big\{(\big\{\{x,y\}\big\}\times {\cal R}^{({\cal F},0)})\cup(\big\{\{x\},\{y\}\big\}\times {\cal R}^{({\cal F},1)})\cup (\big\{\emptyset\big\}\times{\cal R}^{({\cal F},2)}).$$
  As the size of the elements in ${\cal P}_{e_{\ell}}$ depends only on  ${\cal F},$ it is possible to compute ${\bf tm}^{(e_{\ell})}_{\cal F}$ in $\Ocal_{d}(1)$ steps.

  Let $e\in \{e_{r}\}\cup E(T\setminus L(T)),$ and let $e_{1}$ and $e_{2}$ be
  the two other edges of $T$ that share an endpoint with $e$ and where each path from them to $r$ contains $e$. We also set $F_{e}=\big(R_{e_{1}}\cup R_{e_{2}}\big)\setminus R_{e}$.
  For the dynamic programming algorithm, it is enough to describe how to compute ${\bf tm}_{\cal F}^{(e)}$ given ${\bf tm}_{\cal F}^{(e_i)}, i\in\intv{1,2}$.

  For this, given an $e$-pair $(L,{\bf B})\in{\cal P}_{e}$ where ${\bf B}=(B,R,\lambda),$ it is possible to verify  that
  \begin{eqnarray*}
    {\bf tm}^{(e)}_{\cal F}(L,{\bf B}) & = & \min\big\{{\bf tm}_{\cal F}^{(e_{1})}(L_1,{\bf B}_{1})+{\bf tm}_{\cal F}^{(e_{2})}(L_2,{\bf B}_{2})-|L_{1}\cap L_{2}|\mid \\
                                       & & ~~~~~~ \mbox{$(L_{i},{\bf B}_{i})\in{\cal P}_{e_{i}},i\in\intv{1,2}$}, \\
                                       & & ~~~~~~ L_{i}\setminus F_{e}=L\cap R_{e_{i}}, i\in\intv{1,2},\\
                                       & & ~~~~~~ L_{1}\cap R_{e_1} \cap R_{e_2}=L_{2}\cap R_{e_1} \cap R_{e_2}, \mbox{~and~} \\
                                       & & ~~~~~~ {\bf B}={\sf rep}_{\cal F}\Big(\big(({\bf B}_{1}\diamond Z_{1})\odot({\bf B}_{2}\diamond  Z_{2})\big)|_{Z}\Big) \mbox{\ where}\\
                                       & & ~~~~~~~~~~~~~ \mbox{$Z=\rho(R_{e}\setminus L)$ and $Z_{i}=\rho(R_{e_{i}}\setminus L_{i}), i\in\intv{1,2}$}
                                           \big\}.
  \end{eqnarray*}

Note that given ${\bf tm}^{(e_i)}_{\cal F}, i\in\intv{1,2}$ and a pair $(L,{\bf B})\in{\cal P}_{e},$
  the value of ${\bf tm}^{(e)}_{\cal F}(L,{\bf B})$ can be computed by the above formula in  $\Ocal_d(|{\cal P}_{e_1}|\cdot |{\cal P}_{e_2}|)=2^{\Ocal_{d}(w\cdot \log w)}$ steps.  Indeed, by Lemma~\ref{adorning}, the representatives in the sets ${\Rcal}^{({\cal F},t')}$, for $t' \in [0,t]$,  have size $\Ocal_d(w)$, and therefore in the above equation, the boundaried graph $(({\bf B}_{1}\diamond Z_{1})\odot({\bf B}_{2}\diamond  Z_{2})\big)|_{Z}$ has size at most twice the maximum size of a representative, hence $\Ocal_d(w)$ as well. Hence, ${\sf rep}_{\cal F}\Big(\big(({\bf B}_{1}\diamond Z_{1})\odot({\bf B}_{2}\diamond  Z_{2})\big)|_{Z}\Big)$ can be computed by creating a repository (in a preprocessing phase) containing the representative of every graph on at most $\Ocal_d(w)$ vertices, which can be done, by Lemma~\ref{proceeds}, in time $2^{\Ocal_{d}(w\cdot \log w)}$ by using the set  ${\Rcal}^{({\cal F},t)}$ provided by Lemma~\ref{peepshow}.

  As $|{\cal P}_{e}|=2^{\Ocal_{d}(w\cdot \log w)},$ the computation of the function ${\bf tm}^{(e)}_{\cal F}$ requires again $2^{\Ocal_{d}(w\cdot \log w)}$ steps. This means that the whole dynamic programming requires $2^{\Ocal_{d}(w\cdot \log w)}\cdot |E(T)|=2^{\Ocal_{d}(w\cdot \log w)}\cdot \Ocal(|E(G)|)$ steps.  As $|E(G)|=\Ocal(\bw(G)\cdot |V(G)|),$ the claimed running time follows.
\end{proof}


 We conclude this section by observing that the dynamic programming algorithm presented in the proof of Theorem~\ref{oblivion} is  robust, in the sense that it does not explicitly use the fact that the equivalence relation $\equiv^{(\Fcal,t)}$ is defined for the particular {\sl topological minor} containment relation. Indeed, all that the algorithm uses about $\equiv^{(\Fcal,t)}$ is that the size of any representative is $\Ocal_d(t)$ (Lemma~\ref{adorning}) and this allows to conclude that the number of equivalence classes is $2^{\Ocal_{d}(t\cdot \log t)}$ (Lemma~\ref{peepshow}), where $d = {\sf size}(\Fcal)$. Moreover, as discussed at the end of the proof of Theorem~\ref{oblivion}, given a graph of size at most twice the maximum size of a representative, in order to perform the join operation we need to compute (or even precompute) its representative within the claimed asymptotic running time.

Hence, as far as these conditions are fulfilled, the algorithm of Theorem~\ref{oblivion} could also be applied for the \textsc{$\Fcal$-$\frak{C}$-Deletion} problem where $\frak{C}$ is an arbitrary containment relation, and the equivalence relation $\equiv_{\frak{C}}^{(\Fcal,t)}$ is defined according to $\frak{C}$.
This discussion is summarized in the following theorem, which we state here for further reference.

\begin{theorem}\label{thm_DP_for_any_relation}
Let $\frak{C}$ be a graph containment relation, let $\Fcal$ be a proper collection of graphs, let $d = {\sf size}(\Fcal)$, let $t$ be a positive integer, and let $\equiv_{\frak{C}}^{(\Fcal,t)}$ be the equivalence relation on $t$-boundaried graphs defined by $\frak{C}$. Suppose that there exist three functions $f_{{\sf s}} , f_{{\sf n}} , f_{{\sf r}}: \Nbb^2 \to \Nbb$ such that
\begin{itemize}
\item any minimum-sized representative of $\equiv_{\frak{C}}^{(\Fcal,t)}$ has size at most $f_{{\sf s}}(d,t)$,
\item the number of equivalence classes of $\equiv_{\frak{C}}^{(\Fcal,t)}$ is at most  $f_{{\sf n}}(d,t)$, and
\item given a $t$-boundaried graph of size at most $2 f_{{\sf s}}(d,t)$, it is possible to compute a minimum-size representative of its equivalence class of $\equiv_{\frak{C}}^{(\Fcal,t)}$ in time $f_{{\sf r}}(d,t)$.
\end{itemize}
Then the \textsc{$\Fcal$-$\frak{C}$-Deletion} problem can be solved in time $\max\{f_{{\sf n}}(d,t) , f_{{\sf r}}(d,t)  \} ^{\Ocal(1)} \cdot 2^{\Ocal(t)} \cdot n$ on $n$-vertex graphs of treewidth at most $t$.
\end{theorem}

Note that Theorem~\ref{oblivion} is a particular case of Theorem~\ref{thm_DP_for_any_relation} with $\frak{C}$ being the topological minor contaiment relation, $\Fcal$ containing a planar graph, $f_{{\sf s}}(d,t) =\Ocal_d(t)$, and $f_{{\sf n}}(d,t) = f_{{\sf c}}(d,t)  = 2^{\Ocal_{d}(t\cdot \log t)}$.

%

\section{Single-exponential algorithm for planar graphs}
\label{features}



The purpose of this section is to prove Theorem~\ref{deprives}. We do this by proving that, on planar graphs, our dynamic programming approach can be easily modified in order to yield an algorithm  of single-exponential  dependance on \tw.
Later, in Section~\ref{crinkled} we extend the result of this section to graphs of bounded genus.

%
We start with some definitions about planar graphs. We denote the sphere by $\Sigma_{0}$, i.e., $\Sigma_0=\{(x,y,z)\in \Bbb{R}^3 \mid x^2+y^2+z^2=1\}$.
We say that a graph $G$ is {\em  $\Sigma_{0}$-embedded}
if it is
embedded  in $\Sigma_{0}$ without edge crossings.
More generally, given a  disk $\Delta$ (that is a subset of $\Sigma_{0}$ homeomorphic to $\{(x,y)\in \Bbb{R}^2 \mid x^2+y^2\leq 1\}$),
a {\em  $\Delta$-embedded graph} is a graph that is embedded in $\Delta$ without edge crossings and without edges
intersecting its boundary.
To simplify notation,
when we consider a $\Sigma_{0}$-embedded graph or a $\Delta$-embedded graph, we treat it both as the pair $(V(G),E(G))$
and as the set of its points in the embedding.

An {\em $O$-arc} of $\Sigma_{0}$
is a subset of $\Sigma_{0}$ that is homeomorphic to the circle $\{(x,y)\in \Bbb{R}^2 \mid x^2+y^2=1\}$ and an {\em $I$-arc} is a subset of $\Sigma_{0}$ that
is homeomorphic to  the open interval $(0,1)$. Given a $\Sigma_{0}$-embedded graph $G$ and a set $\Lambda\subseteq \Sigma_{0}$, we say that $\Lambda$ is {\em ${G}$-normal} if $\Lambda \cap G\subseteq V(G)$. If an $O$-arc is $G$-normal, then we
call it a \emph{noose} of $G$. Two nooses $N_{1}$ and $N_{2}$ of $G$ are {\em confluent} if one of the two  disks bounded by $N_{1}$  contains $N_{2}$.
%
The \emph{length} of a noose $N$ is the number of  vertices it meets, i.e., $|N|=|V(G)\cap N|$.

\paragraph{Sphere-cut decompositions.}

Let ${\bf G}=(G,R,\lambda)\in{\cal B}^{(t)}$, for some $t\in\Bbb{N}$ and let $\Delta$ be a disk of $\Sigma_{0}$. We say
that $\bound{G}$ is  a {\em $\Delta$-embedded $t$-boundaried graph}
if $G$ is a $\Delta$-embedded graph and $G\cap {\bf bd}(\Delta)=R$, where we use ${\bf bd}(\Delta)$ to denote the boundary of $\Delta$.

A branch-decomposition $(T, \sigma)$ of  a {$\Delta$-embedded boundaried graph} ${\bf G}$ with root $r\in V(T)$ is called a {\em sphere-cut decomposition} (or, in short,  sc-decomposition), if for every edge $e$ of $T$, there exists a noose $N_e$ of $G$, such that\footnote{This definition is slightly different from the original definition given in~\cite{SeymourT94,DornPBF10}, but it can be easily seen that they are  equivalent. Also, note that we consider ${\bf G}$ to be $\Delta$-embedded instead of $\Sigma_0$-embedded; this will be useful to preserve the recursive properties given by the separators of a sc-decomposition.}:
\begin{itemize}
\item The nooses in $\{N_{e}\mid e\in E(T)\}$ are pairwise confluent,
\item $N_{e_{r}}={\bf bd}(\Delta)$,
\item for every $e\in E(T)$, it holds that $G_{e}=\Delta_e\cap G$, where  $\Delta_e$ is the  disk bounded by $N_e$ such that   $\Delta_{e}\subseteq \Delta_{e_r}$,
\end{itemize}
Notice that for every $e\in E(T)$, ${\bf G}_{e}$ can be seen as  a $\Delta_{e}$-embedded $t_{e}$-boundaried graph.
%

The following proposition has been proved in~\cite{SeymourT94,DornPBF10} (for the running time, see~\cite{GuT08}).

\begin{proposition}
  \label{blossoms}
  Let $G$ be a 2-connected $\Sigma$-embedded
  planar graph with $|E(G)|\geq 2$. There exists a sphere-cut decomposition
  of $G$ of width equal to $\bw(G)$. Moreover, a sphere-cut decomposition of $G$ of optimal width
  can be computed in $\Ocal(n^{3})$ steps.
\end{proposition}

%
%
%
\paragraph{Oriented disk-embedded boundaried graphs.}\hspace{-3mm}An  {\em oriented disk-embedded $t$-bounda\-ried graph}
${\bf G}$ (in short, ode-$t$-boundaried graph) is a
quadruple ${\bf G}=(G,\Delta,v,{\sf s})$
where $\Delta$ is a  disk, $G$ is a $\Delta$-embedded graph, $|G\cap {\bf bd}(\Delta)|=t$,
$v\in G\cap {\bf bd}(\Delta)$, and ${\sf s}\in\{+,-\}$. We call ${\sf s}$ {\em orientation} of ${\bf G}$.
We say that  the set $G\cap {\bf bd}(\Delta)$, denoted by $R({\bf G}),$ is the
{\em boundary} of ${\bf G}$.
We use  ${\cal E}^{(t)}_{\circ}$ to denote all ode-$t$-boundaried graphs.
Given a ${\bf G}\in{\cal E}_\circ^{(t)}$, we define $\overline{\bf G}=(G,R,\lambda)$ where $R=R({\bf G})$, $\lambda(v)=1$, and the rest of the vertices of $R$ are indexed
by consecutive numbers in $\{2,\ldots,|R|\}$, following their order on  ${\bf bd}(\Delta)$ by keeping the interior of $\Delta$ on the right or on the left depending on  whether ${\sf s}=+$ or ${\sf s}=-$. Clearly,  $\overline{\bf G}$ is a consecutive $t$-boundaried graph.

%
%
%
%
%

Notice that, given a sphere-cut decomposition $(T,\sigma)$ of a $\Delta$-embedded boundaried graph
${\bf G}$,  for every $e\in E(T)$, $v\in R_e$, and ${\sf s}\in\{+,-\}$,
the quadruple $(G_{e},\Delta_{e},v,{\sf s})$  is an ode-$t$-boundaried graph.
We will use those ode-$t$-boundaried graphs for our dynamic programming algorithm on sphere-cut decompositions.
More precisely, the tables of the dynamic programming will consist
of this type of objects, and note that,  thanks to the topological structure of the separators in a sc-decomposition, they carry all the information that we need in order to glue them together, namely, $\Delta_{e}$, $v$, and ${\sf s}$. We will prove in Lemma~\ref{premises} that their number is single-exponential and this will yield the running time of Theorem~\ref{deprives}.
To this end, we need to adapt the gluing operation for  ode-$t$-boundaried graphs and, based on this, define a suitable equivalence relation from which the tables will be constructed.

Let ${\bf G}_{i}=(G_{i},\Delta_{i},v_{i},{\sf s}_{i}),i\in[2]$
be two graphs in  ${\cal E}^{(t)}_{\circ}$.
%
Notice that  the graph $\overline{\bf G}_{1}\oplus\overline{\bf G}_2$ defines a $\Sigma_0$-embedded graph: just take the union of the points of the embeddings of $G_{1}$ and $G_{2}$ by identifying same index vertices (in case ${\sf s}_{1}= {\sf s}_2$, we redraw one of the graphs by reflecting it outside its disk).
We extend $\oplus$ by defining ${\bf G}_{1}\oplus {\bf G}_{2}=\overline{\bf G}_{1}\oplus\overline{\bf G}_2$.
%
%

%

We define an equivalence relation $\equiv^{(\Fcal,t)}_{\circ}$ on ${\cal E}_{\circ}^{(t)}$ as follows: Given $\bound{G}_1,\bound{G}_2\in{\cal E}_\circ^{(t)},$ we write
$\bound{G}_1 \equiv_{\circ}^{(\Fcal,t)}\bound{G}_2$ if
\begin{eqnarray*}
  \forall \bound{G}\in{\cal E}_{\circ}^{(t)}\ \Fcal \pretp \bound{G}\oplus \bound{G}_1 &  \iff &  \Fcal \pretp\bound{G}\oplus \bound{G}_2.\label{evaluate}
\end{eqnarray*}
%
%
Notice that we can see each ode-$t$-boundaried graph ${\bf G}=(G,\Delta,v,{\sf s})$ as an embedding in $\Delta$ of the boundaried graph $\overline{{\bf G}}$. This implies that, if we ignore the  way combinatorial graphs are embedded, $\equiv^{(\Fcal,t)}_{\circ}$  induces a coarsening of the restriction of $\equiv^{(\Fcal,t)}$ to the combinatorial boundaried graphs in ${\cal E}_{\circ}^{(t)}$. As for every combinatorial planar graph there is a finite number of ways to embed it in $\Sigma_{0}$ (up to topological isomorphism), Lemma~\ref{entering} can also bound the number of equivalence classes of $ \equiv_{\circ}^{(\Fcal,t)}$.
As we did for $\equiv^{({\cal F},t)}$, we set up a set  ${\Rcal}_{\circ}^{({\cal F},t)}$
of representatives of $\equiv_\circ^{({\cal F},t)}$, defined analogously.

\begin{lemma}
  \label{grateful}
  For every $t\in\Bbb{N}$, the number of different
  ode-$t$-boundaried graphs in ${\cal E}_{\circ}^{(t)}$ with at most $\Ocal(t)$ vertices is $2^{\Ocal(t)}$.
\end{lemma}

\begin{proof}
  The proof is based on the well-known fact that there are $2^{\Ocal(n)}$ $\Sigma_{0}$-embedded graphs on $n$ vertices (up to topological isomorphism). This can be easily verified as follows.
  As a consequence of the results of  Tutte~\cite{Tut62} the number of planar triangulations on $n$ vertices is $2^{\Ocal(n)}$
  (see also~\cite{BonichonGHPS06} for more refined bounds). Since planar triangulations are $3$-connected graphs, by Whitney's theorem~\cite{Whi33}, they have a unique embedding in $\Sigma_{0}$.   Taking into account that every $\Sigma_{0}$-embedded graph can be completed into a $\Sigma_{0}$-embedded triangulation by adding at most $3n-6$ edges, in follows that the number of $\Sigma_{0}$-embedded graphs on $n$ vertices is at most $(256/27)^{n+\Ocal(\log n)}  \cdot 2^{3n-6}= 2^{\Ocal(n)}$.

  Given a $\Delta$-embedded graph $G$ and a set $F\subseteq E(G)$, we say that the pair $(G,F)$
  is a {\em partially edge-annotated $\Delta$-embedded graph}. We extend the concept of topological isomorphism to partially edge-annotated $\Delta$-embedded graphs by demanding annotated edges to be mapped to annotated  edges and non-annotated edges to be mapped to non-annotated edges.
  Combining this with the above upper bound for the number of $\Sigma_{0}$-embedded graphs and the fact that there are at most $2^{3t-6}=2^{\Ocal(t)}$ ways to choose $F$, we conclude
  that there are at most $2^{\Ocal(t)}$ different  partially edge-annotated $\Delta$-embedded graphs.

  We correspond each  $(G,\Delta,v,{\sf s})\in {\cal E}_{\circ}^{(t)}$ to the partially edge-annotated $\Delta$-embedded graph
  $(\hat{G},F)$ where $F$ are the edges corresponding to the $G$-normal $I$-arcs of the set
  ${\bf bd}(\Delta)\setminus V(G)$ and $\hat{G}=(V(G),E(G)\cup F)$ (notice that some of the edges in $F$ may already be edges of $G$). Clearly $\hat{G}$ is also a $\Delta$-embedded graph. Therefore,  the number of different  elements in ${\cal E}_{\circ}^{(t)}$ with at most $\Ocal(t)$ vertices is bounded by the number of different  partially edge-annotated $\Delta$-embedded graphs (that is
  $2^{\Ocal(t)}$, as shown above), multiplied by the $k$ possible choices of $v\in R$ and the 2 possible choices of ${\sf s }\in\{+,-\}$. The lemma follows.
\end{proof}

\begin{lemma}
  \label{premises}
  Let $t\in\Bbb{N}$ and ${\cal F}$ be a proper collection containing a subcubic planar graph, where $d={\sf size}({\cal F}),$
  and let ${\Rcal}_{\circ}^{({\cal F},t)}$ be a set of representatives for $\equiv^{(\Fcal,t)}_{\circ}$.
  Then $|{\Rcal}_{\circ}^{({\cal F},t)}|=2^{\Ocal_{d}(t)}$.
  Moreover, there exists an algorithm that given ${\cal F}$ and $t,$
  constructs a set of representatives ${\Rcal}_{\circ}^{({\cal F},t)}$ in $2^{\Ocal_{d}(t)}$ steps.
\end{lemma}

\begin{proof}
  Observe first that Lemma~\ref{deserves} holds for ${\Rcal}_{\circ}^{({\cal F},t)}$, instead of ${\Rcal}_{}^{({\cal F},t)}$, with essentially the same proof.  Indeed, Lemmas~\ref{lem-few-isolated},~\ref{granting}, and~\ref{likewise} are not affected by this change and, as we argued before, there is an analogue of Lemma~\ref{entering} bounding the size of ${\Rcal}_{\circ}^{({\cal F},t)}$.
  The only difference is that now we have that each boundaried graph, say ${\bf X}\in{\cal B}^{(t)}$,  that appears in the proof is now a member of ${\cal E}^{(t)}_{\circ}$ and therefore, during the gluing operations, we replace it  by  $\overline{\bf X}$ that is the boundaried graph corresponding to ${\bf X}$.

  Based on the above observation, Lemma~\ref{adorning} also holds for ${\Rcal}_{\circ}^{({\cal F},t)}$ instead of ${\Rcal}_{}^{({\cal F},t)}$, as it uses the analogue of Lemma~\ref{deserves} as well as Lemma~\ref{granting} and Proposition~\ref{forwards} (that is also not affected by the change). Now combining this version of
  Lemma~\ref{adorning} with Lemma~\ref{grateful}, we conclude that
  $|{\Rcal}_{\circ}^{({\cal F},t)}|=2^{\Ocal_{d}(t)}$.
  The algorithm claimed in the second statement of the lemma is the same as the corresponding
  one in the proof of Lemma~\ref{peepshow}, with the only difference that now, instead of enumerating
  $t$-boundaried graphs, we enumerate  ode-$t$-boundaried graphs that are $2^{\Ocal_{d}(t)}$ many by  Lemma~\ref{grateful}.
  %
\end{proof}

We are now ready to present the proof of Theorem~\ref{deprives}.

\begin{proof}[Proof of Theorem~\ref{deprives}]
  Let $(G',k)$ be an instance of \pbtm\!\!\!.
  First of all, we consider a tree decomposition  $({\cal X},T')$ of $G'$
  whose bags are its blocks and each block is seen as a $1$-boundaried graph  whose boundary is  some cut-vertex of $G$ (this orientation of the blocks can be done by arbitrarily rooting the tree $T'$ of the tree decomposition).  Given this, we can process each one of the blocks by doing  conventional dynamic programming in order to join them, using as tables those of the algorithm given in the proof of  Theorem~\ref{perishes}. This processing will cost $\Ocal_{d}(1)$ steps per block, therefore
  $\Ocal_{d}(n)$ steps in total. In what follows, we explain how to process each one of the blocks. 

  We consider a block ${\bf G}=(G,R,\lambda)$ of $G'$ where $|R|\leq 1$ (if
  $G$ is not the root block of $(T',{\cal X})$, then $|R|=1$, otherwise $|R|=0$). As each block is 2-connected, we can construct, using the algorithm of Proposition~\ref{blossoms}, an optimal  sphere-cut decomposition $(T,\sigma)$ of ${\bf G}$ in $\Ocal(|V(G)|^{3})$ steps.
  Notice that for each $e\in E(T)$, the graph $G$ can be seen as the graph ${\bf G}_{e}\oplus {\bf G}_{e}'$ where ${\bf G}_{e}=(G_{e},\Delta_{e},v,+)$
  and  ${\bf G}_{e}'=(G_{e}',\Delta_{e},v,-)$, where $G_{e}'=G\cap (\Sigma_{0}\setminus {\bf int}(\Delta_{r}))$ and $v$ is the minimum index vertex in $R_{e}$, where ${\bf int}(\Delta_{r})$ denotes the interior of the disk $\Delta_{r}$. Given these conventions, the dynamic programming is the same as the one in the proof of Theorem~\ref{oblivion}. The only difference is that  now we use ${\cal R}^{({\cal F},t)}_\circ$ instead of ${\cal R}^{({\cal F},t)}$. This latter change
  implies, because of Lemma~\ref{premises}, that the tables of the dynamic programming
  for the edge $e\in E(T)$ have $2^{\Ocal_{d}(t_{e})}$ entries. Therefore the algorithm runs in $2^{\Ocal_{d}(\tw)}\cdot n$ steps in total.
\end{proof}

\section{Single-exponential algorithm for graphs embedded in surfaces}
\label{crinkled}

In this section, we give a concise description of how Theorem~\ref{deprives} can be extended to graphs embedded in
surfaces. The idea is again to treat the graphs processed by the dynamic programming of the proof of Theorem~\ref{oblivion}
as embedded graphs. However, when it comes to embeddings
in surfaces, this is more technical to describe formally. For this, we need to provide
some more definitions.

Surfaces are seen as connected compact 2-manifolds without
boundaries. It is known (see e.g.,~\cite{MoharT01}) that any  surface $\Sigma$ can be obtained, up to homeomorphism, by adding ${\bf eg}(\Sigma)$
\emph{crosscaps} to the sphere, where ${\bf eg}(\Sigma)$ is called the \emph{Euler genus} of $\Sigma$.
We say that a graph $G$ is {\em  $\Sigma_{g}$-embedded}
if it is
embedded without crossings in a surface whose Euler genus is $g$.
Theorems~\ref{deprives} and~\ref{cockneys} can be extended as follows.


\begin{theorem}
  \label{desiring}
  If ${\cal F}$ is a proper collection containing a subcubic planar graph, where $d={\sf size}({\cal F}),$ then
  there exists an algorithm that solves  \pbtm
  on $\Sigma_{g}$-embedded graphs in ${2^{\Ocal_{d+g}(\tw)}}\cdot n^3$ steps.
\end{theorem}

\begin{proof}
  The proof is an adaptation of the proof of Theorem~\ref{deprives} to more general surfaces, using some of the tools developed in~\cite{RueST14}.  A {\em noose} of a $\Sigma_{g}$-embedded graph $G$ is any $G$-normal $O$-arc of $\Sigma$. We say that a noose $N$ is {\em non-contractible} if none of the connected components of $G\setminus N$ is homeomorphic to the interior of a disk. We say that a $\Sigma_{g}$-embedded graph is {\em polyhedrally  $\Sigma_{g}$-embedded} if it is $3$-connected and every non-contractible noose of $G$ has size at least $3$.
  Given a graph $G$, a \emph{polyhedral decomposition} of a $\Sigma_{g}$-embedded graph $G$ is a triple $(A,{\cal X},T)$
  where $A\subseteq V(G), |A|=\Ocal(g)$, and $({\cal X},T)$ is a tree decomposition of $G^A:=G\setminus A$ where
  \begin{itemize}
  \item the adhesion of $({\cal X},T)$ is at most $2$,
  \item for every $t\in V(T)$, the torso of $X_{i}$ is polyhedrally $\Sigma_{g_{i}}$-embedded, for some $g_{i}\leq g$.
  \end{itemize}
  In the above definition, the {\em adhesion} of $({\cal X},T)$ is the maximum size of the intersection between two bags of $({\cal X},T)$. Moreover, the {\em torso} of a bag $X_{i}$
  is the graph obtained by $G[X_{i}]$ if we make adjacent each pair of vertices that both belong to the intersection of $X_{i}$ with some other bag.  We also call $A$ {\em apex set} of $G$. We remark that, during the course of this proof,  the presence of $A$ in the notation $G^A$ should be interpreted as  the ``absence'' of $A$  from some more general graph $G$.

  According to~\cite{RueST14}, given a $\Sigma_{g}$-embedded graph $G$, a polyhedral decomposition of $G$ can be  constructed in $\Ocal(n^3)$ steps.
  We now consider such  a polyhedral decomposition of $G$, we root it at the
  union of the apex set $A$ and the intersection of its bag with its parent bag in $T$ and, as we did the beginning of the proof of Theorem~\ref{deprives}, we process the bags of $({\cal X},T)$ in a bottom-up fashion. The only difference is that now, given that we have a way to process each of the the  ``almost polyhedrical'' bags, we
  do conventional dynamic programming, as in Theorem~\ref{perishes}, for tables that have $\Ocal(g)$ instead of at most 1 vertices. This reduces the proof to the case where the input graph, we call it $G$, contains a set $A$ of $\Ocal(g)$ apex vertices such that are $G^A=G\setminus A$ is a polyhedrally $\Sigma_{g'}$-embedded graph for some $g'\leq g$.
  We also see $G^A$ as a $\Sigma_{g}$-embedded  $t$-boundaried graph ${\bf G}^{A}$ whose boundary has size at most 2. To deal with this, we will use
  an analogue of sphere-cut decompositions for graphs that are polyhedrally embedded in higher-genus surfaces, called {\em surface-cut decomposition}, introduced in~\cite{RueST14}. This requires some definitions in order to extend
  the concept of $\Delta$-embedded $t$-boundaried graph to surfaces.\medskip

  Given a $\Sigma_{g}$-embedded graph $G$, any $G$-normal $O$-arc of $\Sigma_{g}$ is called {\em noose} $N$ of $G$.
  We say that a collection ${\cal N}=\{N_{1},\ldots,N_{s}\}$ of $O$-arcs in $\Sigma_{g}$ is a {\em surface-cut} if they are pairwise confluent, $\bigcup_{1\leq i<j\leq s}N_{i}\cap N_{j}$ is finite, and   $\Sigma_{g}\setminus\cupall{\cal N}$ has {\sl exactly two} connected components. We use the notation $C({\cal N})=\cupall{\cal N}$.
  We say that a subset $\Psi$ of $\Sigma_{g}$ is a {\em $g$-disk} if it is the closure of one of the connected components of $\Sigma_g\setminus C({\cal N})$ for some surface-cut collection ${\cal N}$ of $\Sigma_g$. Notice that there are $\Ocal_{g}(1)$ different $g$-disks up to homeomorphism.
  We say that two $g$-disks $\Psi_{1}$ and $\Psi_{2}$ are {\em $g$-complementary}, if $\Psi_1\cup\Psi_2=\Sigma_{g}$ and $\Psi_{1}\cap \Psi_{2}$ is the boundary of both $\Psi_{1}$ and $\Psi_{2}$.
  We also refer to the $g$-disks $\Psi_{1}$ and $\Psi_{2}$ as {\em the two  disks bounded by} ${\cal N}$.
  Given a $g$-disk $\Psi$, we say that a graph $G$ is {\em  $\Psi$-embedded} if $G$ is  embedded in $\Psi$ without edge crossings and without edges intersecting its boundary. 
  Notice that  we can see ${\bf bd}(\Psi)$, i.e.,  the boundary of $\Psi$,  as a graph if we fix the vertex set to be  $R(G):=V(G)\cap {\bf bd}(\Psi)$ (called {\em frontier} vertices); we call this graph {\em frontier}
  of the $\Psi$-embedded graph ${G}$ and we denote it by $F_{G}$.
  We also denote by $F^A_{G}$ the graph obtained by $F_{G}$ after taking its disjoint union with some set $A$ of isolated vertices (called {\em apex} vertices). Given that $a=|A|$, we call $F_{G}^{A}$ {\em $a$-enhanced frontier graph} of the $\Psi$-embedded graph ${G}$.
  We  denote by ${\cal L}^{(t)}_{\Psi,a}$ the set obtained taking all  $a$-enhanced frontier graphs of all $\Psi$-embedded graphs with $t-a$ vertices in their boundary, and renaming their vertices by numbers in $[t]$.
  Moreover, we  update
  ${\cal L}^{(t)}_{\Psi,a}$ by repetitively removing one graph from any pair of isomorphic graphs
  in this set, as long as such a pair exists. Notice that after this relaxation, all graphs in ${\cal L}^{(t)}_{\Psi,a}$ are pairwise non-isomorphic.
  Moreover, it is easy to prove that if $a=\Ocal(g)$,
  then $|{\cal L}^{(t)}_{\Psi,a}|=2^{\Ocal_{g}(t)}$.  Intuitively, the set ${\cal L}^{(t)}_{\Psi,a}$ encodes all  permissible ways the boundaries of two $t$-boundaried graphs can be identified so that such a gluing will result in a $\Sigma_{g}$-embeddable graph, and its cardinality corresponds to the number of automorphisms
  of every frontier graph $F_{G}$ enhanced with $a$ additional apex vertices.
  Later, we will demand that, while gluing boundaried graphs, frontier vertices are identified to frontier vertices and apex vertices are identified to apex vertices.

  Given a $g$-disk $\Psi$, a $t$-boundaried
  graph ${\bf G}=(G,R,\lambda)$ and a set $A\subseteq V(G)$, we say that $({\bf G},A)$ is an  {\em $a$-almost $\Psi$-embedded $t$-boundaried graph} if $A\subseteq R$, $a=|A|$, $G^{A}:=G\setminus A$ is a $\Psi$-embedded graph, $G^A\cap {\bf bd}(\Delta)=R\setminus A$,
  and  $\lambda$ is an isomorphism from $F_{G}^a$
  to the unique graph in ${\cal L}^{(t)}_{\Psi,a}$ that is isomorphic to $F_{G}^A$
  by an isomorphism that maps $A$ to apex vertices.
  We denote by  ${\cal E}_{\Psi,a}^{(t)}$ the set of all $a$-almost
  $\Psi$-embedded $t$-boundaried graphs and by ${\cal E}_{g}^{(t)}$ the union
  of all ${\cal E}_{\Psi,a}^{(t)}$ for all possible $\Psi$ and $a$.
  Given a $({\bf G},A)\in{\cal E}_{\Psi,a}^{(t)}$
  and a $({\bf G}',A')\in{\cal E}_{\Psi',a'}^{(t')}$,
  we say that $({\bf G},A)$ and $({\bf G}',A')$ are {\em complementary} if
  $\Psi'=\tilde{\Psi},$ $t=t'$,  $F_{G}^{A}$ is isomorphic to $F_{G'}^{A'}$,
  and $\lambda(A)=\lambda(A')$. We denote by ${\cal C}({\bf G},A)$
  the  set of  all members of  ${\cal E}_{g}^{(t)}$  that are complementary to $({\bf G},A)$. For simplicity, when we refer to a member of $({\bf G},\emptyset)\in {\cal E}_{\Psi,0}^{(t)}$ we write ${\bf G}$ instead of $({\bf G},\emptyset)$
  and we call ${\bf G}$ {\em $\Psi$-embedded $t$-boundaried graph}.
  \medskip

  The concept of a surface-cut decomposition was introduced in~\cite{RueST14}
  in order to accelerate dynamic programming algorithms in surfaces. We next present this concept, adapted to the terminology
  that we introduced above.
  A branch-decomposition $(T, \sigma)$ of  a  {$\Psi$-embedded $t$-boundaried graph} ${\bf G}^A=(G^A,R^A,\lambda^{A})$ with root $r\in V(T)$ is called a {\em surface-cut decomposition} if, for every edge $e$ of $T$, there exists a surface-cut collection of nooses  ${\cal N}_e$ of the $\Psi$-embedded graph $G^A$, such that:
  \begin{itemize}
  \item The nooses in $\bigcup_{e\in E(T)} {\cal N}_{e}$ are pairwise confluent,
  \item $N_{e_{r}}={\bf bd}(\Psi)$,
  \item for every $e\in E(T)$, it holds that $G_{e}^A=\Psi_e\cap G^A$, where  $\Psi_e$ is the generalized  disk bounded by $N_e$ such that   $\Psi_{e}\subseteq \Psi_{e_r}$,
  \end{itemize}
  Notice that each $G_{e}^A$ is a $\Psi_{e}$-embedded graph and that all above
  conditions reduce to a sphere-cut decomposition in case $a=0$ and ${\bf G}$
  is a $\Delta$-embedded $t$-boundaried graph.
  We  define ${\bf G}_{e}=(G_{e},R_{e},\lambda_{e})$ as a $t_{e}$-boundaried graph where  $G_{e}$ is obtained by $G^{A}_{e}$ after adding to it the vertices in $A$
  and the edges of $G$ that connect vertices in $A$ with vertices in $G^{A}_e$, $R_{e}=R(G_{e}^{A})\cup A$, and $\lambda_{e}=\rho|_{V(G_{e})}$.
  Moreover, we define the {\em complement} of ${\bf G}_{e}$ as  $ \tilde{{\bf G}}_{e}=(\tilde{G}_{e},R_{e},\lambda_{e})$,  where $\tilde{G}_e=G_e\setminus (V(G_{e})\setminus R_{e})$.
  Notice that for every $e\in E(T)$, $({\bf G}_{e},A)\in {\cal E}_{\Psi_e,a}^{(t_e)}$  and $(\tilde{{\bf G}}_{e},A)\in {\cal E}_{\tilde{\Psi},a}^{(t)}$.
  Keep in mind that, for every $e\in E(T)$, $({\bf G}_{e},A)$ and $({\bf G}_{e},A)$ are complementary members of ${\cal E}_{g}^{(t)}$ and  that ${\bf G}_{e}\oplus \tilde{\bf G}_{e}=G$.

  It was proved in~\cite{RueST14} that there is an algorithm
  that, given a polyhedrally $\Psi$-embedded $t$-boundaried
  graph ${\bf G}^A$ where $\bw({\bf G}^A)\leq w$, it returns
  a surface-cut decomposition of ${\bf G}^A$ of width $\Ocal(w)$, in $2^{\Ocal(w)}\cdot n^3$ steps. By adding   the apex set $A$ to all the $t_{e}$-boundaried graphs ${\bf G}_{e}^{A}$ corresponding to this decomposition, we obtain the decomposition on which
  the algorithm of Theorem~\ref{oblivion} now applies. It is important to keep in mind that now we are working with members of ${\cal E}_{g}^{(t)}$, therefore we need to adapt the equivalence relation $\equiv^{{\cal F},t}$ for ${\cal E}_{g}^{(t)}$.
  Then we use the almost-embeddability  property in order to prove that the
  number of equivalence classes -- and therefore the tables of the dynamic programming -- is bounded by a function that is single-exponential in $t_{e}$.

  We define the equivalence relation $\equiv^{(\Fcal,t)}_{\Psi,a}$ on ${\cal E}_{\Psi,a}^{(t)}$ as follows: Given $(\bound{G}_1,A_{1}),(\bound{G}_2,A_{2})\in{\cal E}_{\Psi,a}^{(t)},$ we say that
  $(\bound{G}_1,A_1) \equiv_{\Psi,a}^{(\Fcal,t)}(\bound{G}_2,A_{2})$ if
  \begin{eqnarray*}
    \forall \bound{G}\in{\cal C}({\bf G},A)\ \ \ \Fcal \pretp \bound{G}\oplus \bound{G}_1 &  \iff &  \Fcal \pretp\bound{G}\oplus \bound{G}_2.\label{holidays}
  \end{eqnarray*}
  We also say that
  $(\bound{G}_1,A_1) \equiv_{g}^{(\Fcal,t)}(\bound{G}_2,A_{2})$
  if both $(\bound{G}_1,A_{1}),(\bound{G}_2,A_{2})$ are $a$-almost $\Psi$-embedded $t$-boundaried graphs and $(\bound{G}_1,A_1) \equiv_{\Psi,a}^{(\Fcal,t)}(\bound{G}_2,A_{2})$. Again using Theorem~\ref{entering}, it is easy to derive that the number of equivalence classes of $\equiv_{g}^{({\cal F},t)}$ depends only
  on $g$, $t$, and $d={\sf size}({\cal F})$.  It is now easy to see that Lemma~\ref{premises} can be extended to a set of representatives ${\Rcal}_{g}^{({\cal F},t)}$ of $\equiv_{g}^{({\cal F},t)}$, i.e., that $|{\Rcal}_{g}^{({\cal F},t)}|=2^{\Ocal_{g+d}(t)}$ and that ${\Rcal}_{g}^{({\cal F},t)}$ can be constructed in $2^{\Ocal_{g+d}(t)}$ steps. Indeed, this is a direct consequence of the fact that an analogue of Lemma~\ref{grateful} can be proved using that $|{\cal L}^{(t)}_{\Psi,a}|=2^{\Ocal_{g}(t)}$ (for $a=\Ocal(g)$) and the fact that there are $2^{\Ocal_{g}(t)}$ $\Sigma_{g}$-embedded graphs on $t$ vertices (this  follows from the bounds on planar graphs by a standard planarization argument cutting along non-contractible cycles).
  Now, by replacing ${\Rcal}_{}^{({\cal F},t)}$ by ${\Rcal}_{g}^{({\cal F},t)}$ in the tables of the dynamic programming algorithm of Theorem~\ref{oblivion}, we can  derive that, for $\Sigma_{g}$-embedded graphs,  the algorithm runs in  $2^{\Ocal_{d+g}(\tw)}\cdot n$ steps, as claimed.
\end{proof}

Again, by Lemma~\ref{sickness} and Observation~\ref{stuffing} we obtain the following counterpart of Theorem~\ref{desiring} for the  minor version.

\begin{theorem}
  \label{cleaning}
  If ${\cal F}$ is a proper collection containing a planar graph, where $d={\sf size}({\cal F}),$ then
  there exists an algorithm that solves \pbm
  on $\Sigma_{g}$-embedded graphs in ${2^{\Ocal_{d+g}(\tw)}}\cdot n^3$ steps.
\end{theorem}



\section{Conclusions and further research}
\label{upheaval}

We presented parameterized algorithms for \textsc{$\Fcal$-M-Deletion} and  \textsc{$\Fcal$-TM-Deletion} taking as  parameter the treewidth of the input graph. These algorithms are complemented by single-exponential algorithms and lower bounds presented in~\cite{monster2,monster3}, and the (optimal) algorithms presented in~\cite{BasteST20-SODA,SODA-arXiv}.

The ultimate goal is to establish the tight complexity of \textsc{$\Fcal$-M-Deletion} and \textsc{$\Fcal$-TM-Deletion} for all collections $\Fcal$.  Recently, the authors made a significant step in this direction~\cite{BasteST20-SODA,SODA-arXiv}, by providing an algorithm to solve \textsc{$\Fcal$-M-Deletion} in time $\Ostar(2^{\Ocal(\tw \cdot \log \tw)})$ for {\sl every} collection $\Fcal$. This algorithm uses, as a black box, the algorithm of Section~\ref{dramatic} (as stated in Theorem~\ref{thm_DP_for_any_relation}), as well as other ingredients such as Bidimensionality~\cite{FominDHT16}, the irrelevant vertex technique~\cite{RobertsonS12,RobertsonS95b}, and recent results about rerouting paths on flat structures~\cite{GolovachST20}. Note that this result vastly generalizes the ones of Jansen et al.~\cite{JansenLS14} and Kociumaka and Pilipczuk~\cite{KociumakaP17}, running in time $\Ostar(2^{\Ocal(\tw \cdot \log \tw)})$, for the problems of deleting a minimum number of vertices to obtain a planar graph and a graph of Euler genus at most $g$, respectively. Combined with the lower bounds presented in~\cite{monster3}, the algorithm in~\cite{BasteST20-SODA} settles completely the complexity of \textsc{$\Fcal$-M-Deletion} when $\Fcal$ consists of a single connected graph. Determining the tight complexity, conceivably either $\Ostar(2^{\Ocal(\tw)})$ or $\Ostar(2^{\Ocal(\tw \cdot \log \tw)})$, when $\Fcal$ contains more than one graph, or when the graph(s) in $\Fcal$ may be disconnected, remains still open.

On the other hand, we still do not have such a general algorithm running in time $\Ostar(2^{\Ocal(\tw \cdot \log \tw)})$ for the \textsc{$\Fcal$-TM-Deletion} problem, the fastest one being the algorithm of Theorem~\ref{perishes}. In particular, we do not know whether there exists some $\Fcal$ for which there is a lower bound of, say, $2^{o(\tw^2)} \cdot n^{\Ocal(1)}$ for \textsc{$\Fcal$-TM-Deletion}.

We presented single-exponential algorithms for \textsc{$\Fcal$-Deletion} when the input graph is planar or, more generally, embedded in a fixed surface. In both cases, the key tool is a special type of branch decomposition with nice topological properties. It seems plausible that this result could be extended to input graphs excluding a fixed graph $H$ as a minor, by using the so-called \emph{$H$-minor-free cut decompositions} introduced by Ru\'e et al.~\cite{RueST12-COCOON}.

In the last years, the \textsc{$\Fcal$-M-Deletion} problem has been extensively studied in the literature taking as the parameter the size of the solution~\cite{KimLPRRSS16line,FominLMS12,JoretPSST14,KPP15,JansenLS14,SST20-ICALP-arXiv}. In all these papers, {\sf FPT}-algorithms parameterized by treewidth play a fundamental role. The results that we presented in this paper have already been used in~\cite{BasteST20-SODA}, which in turn has been strongly used in the algorithms in~\cite{SST20-ICALP-arXiv}.

Our results have also interesting consequences by applying the Bidimensionality framework~\cite{FominDHT16,DemaineH07-CJ,DemaineFHT05jacm}, as we proceed to discuss. Let $\Fcal$ be a collection containing a planar graph and consider the  \textsc{$\Fcal$-M-Deletion} problem parameterized by the solution size $k$, restricted to input graphs $G$ that exclude some fixed graph $H$ as a minor.
The linearity in terms of treewidth of the size of a largest grid in an $H$-minor-free graph~\cite{DemaineH05a} implies that positive instances of  \textsc{$\Fcal$-M-Deletion} have treewidth $\Ocal_{\Fcal,H}(\sqrt{k})$. Indeed, otherwise, for any solution $S \subseteq V(G)$ of size at most $k$, $G \setminus S$ would contain as a minor  a large enough grid, as a function of $\Fcal$, which would contain a planar graph in $\Fcal$ as a minor, contradicting the fact that $S$ is a solution. Therefore, when $\Fcal$ contains a planar graph, Theorem~\ref{headings} yields an algorithm to solve \textsc{$\Fcal$-M-Deletion} in time ${2^{\Ocal_{\Fcal,H}(\sqrt{k}\cdot \log k)}}\cdot n$ when the input graph is $H$-minor-free, and, by Theorem~\ref{cleaning}, in time ${2^{\Ocal_{\Fcal,g}(\sqrt{k})}}\cdot n$ when the input graph has genus at most $g$. Plausibly, using the results in~\cite{RueST12-COCOON}, the running time ${2^{\Ocal_{\Fcal,g}(\sqrt{k})}}\cdot n$ could be achieved also for $H$-minor-free graphs. To be best of our knowledge, subexponential algorithms for \textsc{$\Fcal$-M-Deletion} on these classes of graphs were not known before.

\vspace{.3cm}

\noindent \textbf{Acknowledgements}. We would like to thank the referees of the two conference versions containing some of the results of this article, and those of the current version,  for helpful remarks that improved the presentation of the manuscript, in particular for pointing out the fact that our results yield subexponential algorithms parameterized by the size of the solution.


\bibliographystyle{abbrv}
\bibliography{Biblio-Fdeletion}

\end{document}